\documentclass[10pt]{olplainarticle}

\title{\Large{Superstructure Optimization with Embedded Neural Networks for Sustainable Aviation Fuel Production}}

\author[a]{\footnotesize{Alexander Klimek}}
\author[bc]{\footnotesize{Christoph Plate}}
\author[bc]{\footnotesize{Sebastian Sager}}
\author[ad]{\footnotesize{Kai Sundmacher}}
\author[*a]{\footnotesize{Caroline Ganzer}}
\affil[a]{\footnotesize{Max Planck Institute for Dynamics of Complex Technical Systems, Department of Process Systems Engineering, Sandtorstr. 1, 39106 Magdeburg, Germany.}}
\affil[b]{\footnotesize{Max Planck Institute for Dynamics of Complex Technical Systems, Mathematical Optimization and Machine Learning Group, Sandtorstr. 1, 39106 Magdeburg, Germany.}}
\affil[c]{\footnotesize{Otto von Guericke University, Chair of Mathematical Algorithmic Optimization, Universitätsplatz 2, 39106 Magdeburg, Germany.}}
\affil[d]{\footnotesize{Otto von Guericke University, Chair for Process Systems Engineering, Universitätsplatz 2, 39106 Magdeburg, Germany.}}
\affil[*]{\footnotesize{corresponding author; E-mail: cganzer@mpi-magdeburg.mpg.de}}


\begin{abstract}
This study presents a multi-objective optimization framework for sustainable aviation fuel (SAF) production, integrating artificial neural networks (ANNs) within a mixed-integer quadratically constrained programming (MIQCP) formulation. By embedding data-driven surrogate models into the mathematical optimization structure, the proposed methodology addresses key limitations of conventional superstructure-based approaches, enabling simultaneous optimization of discrete process choices and continuous operating parameters. The framework captures variable input and output stream compositions, facilitating the joint optimization of target product composition and system design. Application to Fischer-Tropsch (FT) kerosene production demonstrates that cost-minimizing configurations under unconstrained \ce{CO2} emissions are dominated by the fossil-based autothermal reforming (ATR) route. Imposing carbon emission constraints necessitates the integration of biomass gasification and direct air capture coupled with carbon sequestration (DAC-CS), resulting in substantially reduced net emissions but higher production costs. At the zero-emission limit, hybrid configurations combining ATR and biomass gasification achieve the lowest costs (\textapprox\SI{2.38}{\$/kg\textsubscript{kerosene}}), followed closely by biomass gasification-only (\textapprox\SI{2.43}{\$/kg}), both of which outperform the ATR-only pathway with DAC-CS (\textapprox \SI{2.65}{\$/kg}). In contrast, DAC-only systems relying exclusively on atmospheric \ce{CO2} and water electrolysis are prohibitively expensive (\textapprox\SI{10.8}{\$/kg}). The results highlight the critical role of the embedded ANNs: optimal process conditions, such as FT reactor pressure and gasification temperature, adapt to changing circumstances, consistently outperforming fixed setups and achieving up to \SI{20}{\%} cost savings.
\end{abstract}

\begin{document}

\flushbottom
\maketitle
\thispagestyle{empty}

\renewcommand{\sectionautorefname}{Section} 
\renewcommand{\subsectionautorefname}{Subsection} 
\renewcommand{\subsubsectionautorefname}{Subsubsection} 
\renewcommand{\equationautorefname}{Eq.} 

\section{Introduction}
The defossilization of aviation represents a formidable challenge in the global transition toward net zero greenhouse gas (GHG) emissions. In 2019, civil aviation was responsible for over 900 million metric tons of \ce{CO2}-equivalent emissions, accounting for approximately \SI{10}{\%} of global transportation-related emissions \cite{Vardon.2022}. Forecasts anticipate a four- to six-fold increase in air traffic over the coming decades, driven by population growth, urbanization, and globalization \cite{Bauen.2020, GonzalezGaray.2022}. Whilst \ce{H2}-fueled aircrafts and electrification may present viable options for short- and medium-haul air traffic and niche applications, it is expected that long-chain liquid sustainable aviation fuels (SAFs) are crucial at large scale \cite{FreireOrdonez.2022}. 

SAFs are hydrocarbon-based fuels synthesized from renewable carbon (\ce{C}) and \ce{H2} sources \emph{via} certified thermochemical or biochemical conversion pathways. Unlike alternative propulsion technologies, SAFs are compatible with existing aircraft engines, fuel handling systems, and airport infrastructure. As such, they can serve as drop-in replacements for conventional jet fuels (\emph{e.g.}\@, Jet-A1), facilitating a smoother and less disruptive transition to low-carbon aviation. The American Society for Testing and Materials (ASTM) International has codified the production and compositional criteria for SAFs under Standard D7566, with multiple conversion pathways such as Fischer–Tropsch Synthetic Paraffinic Kerosene (FT-SPK), Hydroprocessed Esters and Fatty Acids (HEFA-SPK), and Alcohol-to-Jet Synthetic Paraffinic Kerosene (ATJ-SPK) having already received certification for commercial deployment \cite{ASTMInternational., Bauen.2020, Holladay.2020}.

However, the rational design of economically and environmentally optimal SAF production pathways remains a multi-dimensional optimization problem. This complexity arises from the wide array of feasible raw material sources, including biomass, captured \ce{CO2}, and renewable \ce{H2}, as well as the multiplicity of intermediate chemistries (\emph{e.g.}\@, syngas, methanol, FT liquids), conversion process technologies, and energy integration schemes \cite{Bube.2025}. These process network design problems are further complicated by nonlinear thermodynamic relationships, discrete design choices, and sustainability constraints such as \ce{CO2} emissions and land use. Identifying viable SAF production configurations requires a systematic modeling approach that accounts for many process alternatives, their interdependencies, and interactions among material and energy flows as well as emissions \cite{GonzalezGaray.2022}. 

Recent techno-economic analyses have assessed various power-to-liquid fuel production routes, revealing significant cost disparities and efficiency trade-offs. For instance, Salem \cite{Salem.2023} emphasizes the wide cost range of methanol-to-jet (MtJ) pathways (\SI{650}{\euro} to \SI{4630}{\euro} per ton) primarily driven by the price of \ce{H2} and \ce{CO2}. Sacchi \emph{et al.}\@ \cite{Sacchi.2023} present a time-dependent life-cycle assessment comparing fossil jet fuel with DAC-CS offsets to synthetic jet fuel from DAC and electrolysis. They show that climate outcomes depend strongly on the carbon footprint of electricity, with synthetic fuels reducing \ce{CO2} storage needs if powered by low-carbon electricity. Do \emph{et al.}\@ \cite{Do.2022} develop a techno-economic framework to evaluate \ce{CO2}-to-fuel pathways, including aviation fuels, using process simulation and optimization to assess various technologies and scenarios. Meanwhile, Albrecht \emph{et al.}\@ \cite{Albrecht.2017} demonstrate that the viability of different SAF pathways depends heavily on electricity prices, with biomass-to-liquid favored under higher costs of electricity. Comparative studies by Bube \emph{et al.}\@ \cite{Bube.2024, Bube.2025} and Voß \emph{et al.}\@ \cite{Vo.2024} indicate that MtJ offers higher kerosene selectivity and lower energy input, while hybrid routes based on biomass and electrolytic \ce{H2} achieve higher C efficiency at a premium cost. Eyberg \emph{et al.}\@ \cite{Eyberg.2024} underline the persistent economic gap with fossil jet fuel, suggesting that policy support and technological advancements will be pivotal.

A growing body of research has applied optimization-based approaches to the synthesis and integration of SAF production pathways, adopting a variety of system boundaries and modeling strategies. Gonzalez-Garay \emph{et al.}\@ \cite{GonzalezGaray.2022} formulate a superstructure optimization of solar-powered SAF production based on Aspen Plus\textsuperscript{\textregistered} \cite{AspenPlu.24.08.2023} simulations, using fixed operating conditions and lumped component modeling. Demirhan \emph{et al.}\@ \cite{Demirhan.2021} develop a multi-scale optimization framework for fuels, chemicals, and power generation, incorporating multiple \ce{H2} production routes and policy incentives. Martín and Grossmann \cite{Martin.2018} and Zhang \emph{et al.}\@ \cite{Zhang.2018} analyze renewable fuels and power integration in Spain using linear models with temporal discretization. Kenkel \emph{et al.}\@ \cite{Kenkel.2021} integrate Power-to-Jet and Biomass-to-Jet production in a refinery superstructure including algae-based feedstocks, while Niziolek \emph{et al.}\@ \cite{Niziolek.2017} examine municipal solid waste conversion to fuels and chemicals \emph{via} a broad reaction network. Finally, Wang \emph{et al.}\@ \cite{Wang.2013} propose a multiobjective Mixed-Integer Nonlinear Programming (MINLP) model for biomass gasification-based hydrocarbon biorefineries, integrating economic and environmental criteria. Across these studies, simplifications are common in the treatment of component mixtures, operating conditions, and carbon management. Most approaches rely on lumped components and linear process models, fixed operating parameters, and predetermined stream compositions. Limited consideration of biomass, atmospheric carbon, and carbon capture further restricts flexibility in design. For these reasons, there is an urgent need for advanced modeling frameworks that incorporate nonlinear process behavior, detailed mixture representations, variable operating conditions, broader feedstock options, and emissions management strategies to more fully capture the trade-offs in SAF production.

To support such advanced formulations, research on superstructure optimization has explored multiple representations of processes and streams. Farkas \emph{et al.}\@ introduce the R-graph representation as an extension of the state-equipment network, in which nodes correspond to unit-operation ports and arcs represent feasible flow connections between them, enabling a compact encoding of alternative separation configurations while exploiting graph-theoretic properties \cite{Farkas.2005}. Building on these concepts, Wu \emph{et al.}\@ propose a framework that decomposes a process into units, inlet and outlet ports, and conditioning streams. Sources and sinks supply and collect material, ports act as generalized mixers and splitters, and conditioning streams handle temperature and pressure adjustments \cite{Wu.2016}. Mencarelli \emph{et al.}\@ provide a comprehensive review of such approaches, comparing state-task networks, P-graphs, R-graphs, port-based graphs, and modular building-block grids in terms of suitability for various classes of process synthesis problems \cite{Mencarelli.2020}.

Within this general superstructure landscape, several contributions incorporate mixtures to connect process decisions with fuel properties. Restrepo-Flórez and Maravelias \cite{RestrepoFlorez.2021} formulate a superstructure for upgrading ethanol to advanced fuels in which individual product species are grouped into surrogate fuel components that are blended to meet viscosity, octane, and distillation constraints. Their mixed-integer models balance flowsheet decisions with composition-dependent fuel properties in a unified framework. Along similar lines, Daoutidis and colleagues develop pathway superstructures for fuel production from biomass-derived intermediates. Alternative reaction routes are optimized subject to energy, carbon, and fuel-quality considerations. Mixture properties such as energy density and oxygen content guide the selection of products and processing routes \cite{Marvin.2013, Rangarajan.2014}. Reaction Network Flux Analysis and Process Network Flux Analysis have been combined with mixture property models to screen biofuel routes based on both the feasibility of the fuel synthesis and target fuel attributes, without resorting to detailed flowsheet simulations \cite{Dahmen.2016, Dahmen.2017, Hechinger.2010, Konig.2020, Panofen.2025, Ulonska.2016, Voll.2011}. In more detailed process superstructures, multi-component stream compositions are tracked through separators and blending units, and nonlinear mixture constraints are imposed at pool or product nodes \cite{Galan.1998, Karuppiah.2006, Ahmetovic.2011}. These formulations with mixtures illustrate how embedding fuel property and blending models directly into the optimization problem is essential for designing processes that not only maximize yield or profit, but also deliver on-spec fuels.

Surrogate modeling complements these structural advances by enabling tractable optimization of the resulting large and nonlinear models. As summarized by McBride and Sundmacher \cite{McBride.2019}, commonly used surrogate models include polynomials, Gaussian processes (Kriging), ANNs, and radial basis functions. While Kriging dominates surrogate-based optimization studies, ANNs are increasingly used for high-dimensional, nonlinear problems. Surrogates are typically trained \emph{via} sequential data generation from simulation platforms such as Aspen Plus\textsuperscript{\textregistered} \cite{AspenPlu.24.08.2023}. Several studies have explored the integration of surrogate models into superstructure or flowsheet optimization. Fahmi and Cremaschi \cite{Fahmi.2012} embed ANNs trained on Aspen Plus\textsuperscript{\textregistered} \cite{AspenPlu.24.08.2023} simulation data in a superstructure to optimize a biodiesel production for describing nonlinear process behaviors with regard to conversion and utility demand. Their disjunctive programming formulation leads to a MINLP solved efficiently in GAMS. Similarly, Henao and Maravelias \cite{Henao.2011} propose a generalized framework for surrogate-based superstructure optimization, replacing detailed unit models with ANNs. They emphasize the importance of sampling strategy, surrogate architecture, and variable selection, demonstrating optimization of \ce{CO2} capture, distillation, and reaction systems using full-space ANN formulations in GAMS. To address scalability, Schweidtmann and Mitsos \cite{Schweidtmann.2018} develop a deterministic global optimization method using McCormick relaxations within reduced-space ANN formulations. This reduces optimization complexity significantly by avoiding explicit representation of ANN internals, a concept further elaborated by Schweidtmann \emph{et al.}\@ \cite{SchweidtmannPardolos.2022}, who distinguish between full-space and reduced-space formulations for embedding machine learning models in optimization tasks. These studies show that deep ANNs, while accurate, increase computational cost unless an appropriate problem reformulation is applied. Granacher \emph{et al.}\@ \cite{Granacher.2021} integrate ANNs, random forests, and Gaussian process surrogates within an active learning loop to optimize biomass-to-liquid process flowsheets. Complementary studies include Pedrozo \emph{et al.}\@ \cite{Pedrozo.2020}, who used piecewise linear surrogates refined iteratively for ethylene production technology screening, and Hao \emph{et al.}\@ \cite{Hao.2022}, who trained ANNs to replace carbon capture and utilization subsystem simulations, leading to faster multi-objective optimization. For a comprehensive literature overview regarding the mathematical complexity of embedding surrogate models in optimization problems, the reader is referred to Plate \emph{et al.}\@ \cite{Plate.252025}. In summary, the literature reveals a growing maturity in surrogate-assisted optimization frameworks. Still, challenges persist in ensuring optimization feasibility outside training domains, choosing appropriate surrogate architectures, and balancing model accuracy with computational tractability, particularly in complex flowsheets and SAF-related systems.

Superstructure optimization provides a rigorous mathematical framework for the identification of optimal SAF production routes from both an economic and environmental perspective. Potentially feasible processes and interconnections can be embedded into a comprehensive network representation. The optimal design is then determined by solving an optimization problem, typically a Mixed-Integer Linear or Nonlinear Programming (MILP/MINLP) problem. However, conventional formulations often rely on oversimplified assumptions, such as neglecting composition-dependent behavior and linearizing inherently nonlinear phenomena \cite{Demirhan.2021, Ganzer.2020, GonzalezGaray.2022, Niziolek.2017, Svitnic.2022}. These approximations are particularly limiting for SAF synthesis, where product quality (defined by distributions of hydrocarbon chain lengths) and conversions are strongly composition-dependent \cite{Holladay.2020}.

To address these limitations, we propose an advanced superstructure optimization framework. We explicitly model mixtures within the superstructure by including mass fractions as decision variables. Our formulation yields a Mixed-Integer Quadratically Constrained Programming (MIQCP) problem. Importantly, this structure remains compatible with global optimization using modern solvers such as Gurobi \cite{GurobiOptimization.05.10.2023}. A key innovation of this work is the integration of ANNs as surrogate models within the superstructure optimization. These embedded ANNs are trained on flowsheet simulations and capture the nonlinear dependence of characteristic process variables, such as selectivity, conversion, and energy demand, as a function of operating conditions and inlet compositions. The primary contributions of this study can be summarized as follows:
\begin{itemize}[noitemsep, topsep=0pt]
\item development of an advanced MIQCP-based superstructure formulation that incorporates stream compositions and enables the optimization of mass fractions across processes,
\item embedding of ANNs into the optimization problem to model nonlinear process behavior and include operating conditions as decision variables,
\item application of the optimization framework to FT-based SAF production,
\item demonstration that allowing greater degrees of freedom at the unit operation level, enabled through neural network surrogates, leads to superior overall system performance in terms of cost and \ce{CO2} emission intensity.
\end{itemize}

The remainder of this paper is structured as follows. We introduce the complete MIQCP superstructure optimization formulation in \autoref{Section: Superstructure optimization with mixtures/model description}. The specific components and processes involved, including a description of detailed flowsheet simulations, for applying our optimization formulation are described in \autoref{Section: Data set for Fischer-Tropsch sustainable aviation fuel}. \autoref{Section: Process representation via neural networks} provides a comprehensive description of the training of ANNs, and their subsequent integration into the overarching superstructure optimization framework. \autoref{Section: Results} presents the results of the study, and the last section concludes with key findings and avenues for future work.

\section{Superstructure optimization with mixtures} \label{Section: Superstructure optimization with mixtures/model description}
\begin{figure}[h]
\centering
  \includegraphics[trim=2.1cm 1.25cm 2.22cm 4.6cm, clip, width=\columnwidth]{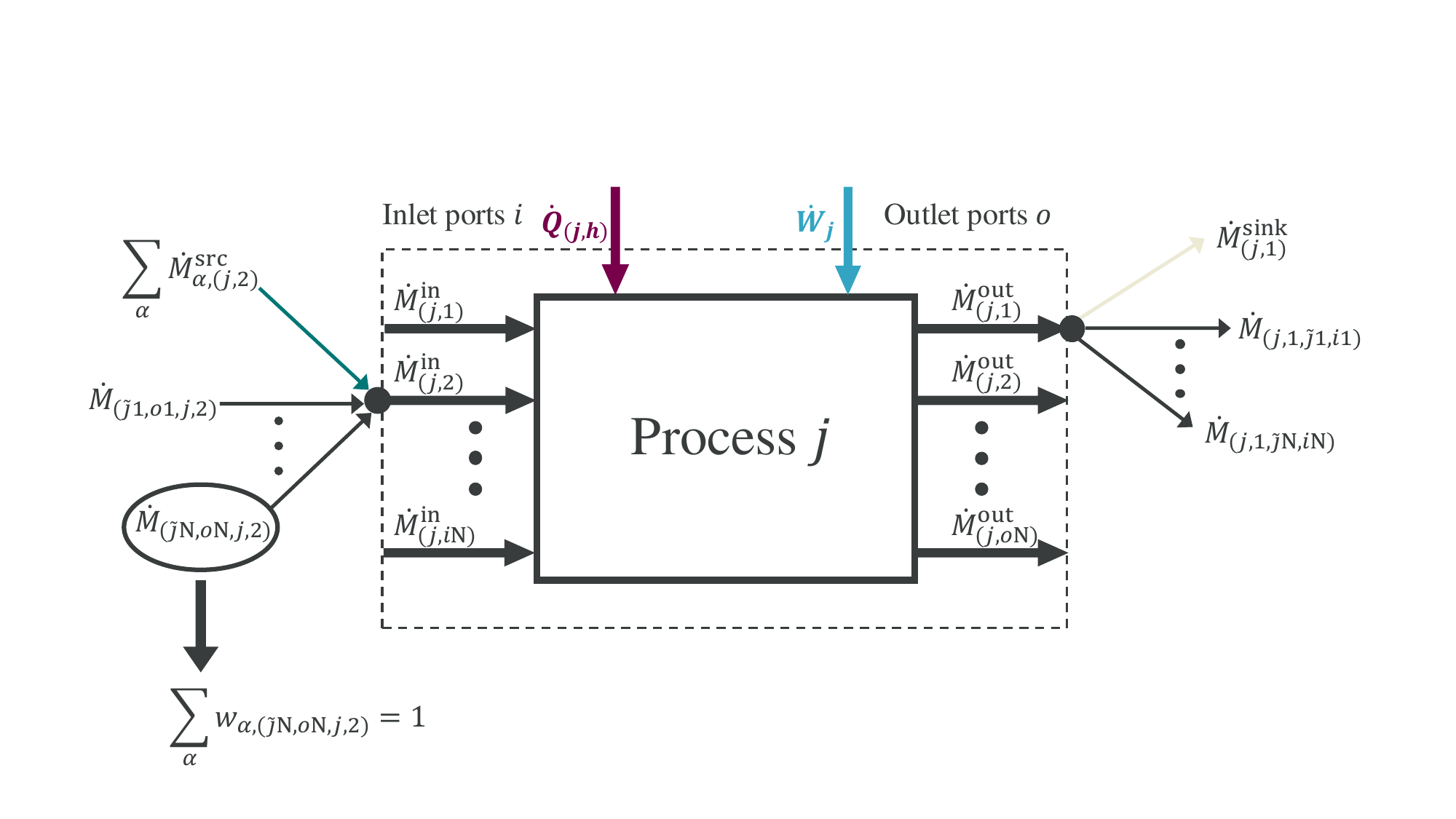}
\caption{Generic model of a chemical conversion process within the superstructure network. Each process is characterized by inlet and outlet ports through which mass flows can enter and exit the process. The general numbers of inlet or outlet ports are denoted by $\mathrm{iN}$ and $\mathrm{oN}$, respectively. A process can contain several unit operations (\emph{e.g.}\@, compression, heating/cooling, reaction, separation). The process boundaries are indicated with solid lines, while the boundaries for balancing the inlet and outlet ports are shown with dashed lines.}
\label{Fig.: General process modeling.}
\end{figure}

\autoref{Fig.: General process modeling.} presents the generalized formulation adopted for modeling all processes within the superstructure network through the explicit definition of multiple inlet and outlet ports. This framework facilitates the integration of surrogate models, specifically ANNs, which are employed to capture the complex, nonlinear input–output relationships characteristic of steady-state process flowsheets.
These ANNs are capable of describing the transformation of multiple chemical species entering the processes \emph{via} distinct inlets and exiting through multiple outlet streams. Our framework is thus an extension of the underlying approaches by Ganzer and Mac Dowell \cite{Ganzer.2020} as well as Svitnič and Sundmacher \cite{Svitnic.2022}, and incorporates concepts from Ahmetović and Grossmann \cite{Ahmetovic.2011} and Wu \emph{et al.}\@ \cite{Wu.2016}. Key features of our framework include:
\begin{itemize}[noitemsep, topsep=0pt]
\item explicit consideration of multiple process inlet and outlet ports, enabling flexible representation of the connections between processes,
\item representing streams as mixtures allowing for the optimization of compositions,
\item integration of ANN-based surrogate models, thereby incorporating nonlinear process behaviors while maintaining computational tractability.
\end{itemize}
The mathematical expressions used to formulate the MIQCP model are explained in detail below, with processes denoted by the index $j$, components by $\alpha$, and inlet/outlet ports by $i$/$o$.

\subsection{Mass balance constraints}
We establish generalized total mass balance equations for each inlet port $(j,i)$ associated with every process in the superstructure network, expressed as:
\begin{align}
\dot{M}_{(j,i)}^{\mathrm{in}} = \sum_{(\tilde{j},o):(\tilde{j},o,j,i) \in \mathcal{J}^{\mathrm{OI}}} \dot{M}_{(\tilde{j},o,j,i)} + \sum_{\alpha \in \mathcal{C}} \dot{M}_{\alpha,(j,i)}^{\mathrm{src}} \quad \forall (j,i) \in \mathcal{J}^{\mathrm{I}},
\end{align}
as depicted in \autoref{Fig.: General process modeling.}. Index pairs in parentheses represent specific pairings that are predefined by the set definitions. In our formulation, $\dot{M}_{(j,i)}^{\mathrm{in}}$ denotes the total mass flow rate entering inlet port $i$ of process $j$. The term $\dot{M}_{(\tilde{j},o,j,i)}$ refers to the mass flow rate transferred from outlet port $o$ of process $\tilde{j}$ to the specified inlet port $(j,i)$. External source contributions (\emph{i.e.}\@, mass inputs entering the system boundary) are represented by $\dot{M}_{\alpha,(j,i)}^{\mathrm{src}}$, where $\alpha$ indexes the set of all species $\mathcal{C}$. The set $\mathcal{J}^{\mathrm{OI}}$ encompasses all admissible outlet-to-inlet port connections across the superstructure network, while $\mathcal{J}^{\mathrm{I}}$ defines the collection of all process inlet ports. A comprehensive overview of all sets involved in the formulation is provided in Subsection A.2.\@ of the supplementary information (SI). Component-wise (partial) mass balances are additionally imposed for each process inlet port $(j,i)$, formulated as
\begin{align}
w_{\alpha,(j,i)}^{\mathrm{in}} \dot{M}_{(j,i)}^{\mathrm{in}} &= \sum_{(\tilde{j},o):(\tilde{j},o,j,i) \in \mathcal{J}^{\mathrm{OI}}} w_{\alpha,(\tilde{j},o,j,i)} \dot{M}_{(\tilde{j},o,j,i)} \nonumber \\
&\quad + \dot{M}_{\alpha,(j,i)}^{\mathrm{src}} \quad \forall (j,i) \in \mathcal{J}^{\mathrm{I}}, \; \alpha \in \mathcal{C}, \label{EQ: 2}
\end{align}
where $w_{\alpha}$ denotes the mass fraction of species $\alpha$ within the associated total mass stream. The mass inflow of all non-source components from outside the system boundaries is set to zero:
\begin{align}
\dot{M}_{\alpha,(j,i)}^{\mathrm{src}} = 0 \quad \forall (j,i) \in \mathcal{J}^{\mathrm{I}}, \; \alpha \in \mathcal{C} \setminus \mathcal{C}^{\mathrm{src}},
\end{align}
where $\mathcal{C}^{\mathrm{src}} \subseteq \mathcal{C}$ is the subset of components permitted to enter the system as external source streams. Furthermore, mass balance constraints are formulated to define the aggregate inlet stream characteristics for each process across all associated inlet ports. The total and partial mass flow rate entering process $j$ is given by
\begin{align}
\dot{M}_{j}^{\mathrm{in,tot}} &= \sum_{i:(j,i) \in \mathcal{J}^{\mathrm{I}}} \dot{M}_{(j,i)}^{\mathrm{in}} \quad \forall j \in \mathcal{J}, \\
w_{\alpha,j}^{\mathrm{in,tot}} \dot{M}_{j}^{\mathrm{in,tot}} &= \sum_{i:(j,i) \in \mathcal{J}^{\mathrm{I}}} w_{\alpha,(j,i)}^{\mathrm{in}} \dot{M}_{(j,i)}^{\mathrm{in}} \quad \forall j \in \mathcal{J}, \; \alpha \in \mathcal{C}, \label{EQ: 5}
\end{align}
where $\mathcal{J}$ denotes the set of all processes in the superstructure. \autoref{EQ: 2} and \autoref{EQ: 5} can be regarded as mixing constraints.

An analogous formulation is employed for total and component-specific mass balances at each outlet port $(j,o)$ in the process network. The total mass flow exiting an outlet port is expressed as
\begin{align}
\dot{M}_{(j,o)}^{\mathrm{out}} = \sum_{(\tilde{j},i):(j,o,\tilde{j},i) \in \mathcal{J}^{\mathrm{OI}}} \dot{M}_{(j,o,\tilde{j},i)} + \dot{M}_{(j,o)}^{\mathrm{sink}} \quad \forall (j,o) \in \mathcal{J}^{\mathrm{O}},
\end{align}
where $\dot{M}_{(j,o)}^{\mathrm{sink}}$ represents the terminal outflow exiting the superstructure boundaries. The corresponding partial mass balances for each chemical species $\alpha \in \mathcal{C}$ are formulated as:
\begin{align}
w_{\alpha,(j,o)}^{\mathrm{out}} \dot{M}_{(j,o)}^{\mathrm{out}} &= \sum_{(\tilde{j},i):(j,o,\tilde{j},i) \in \mathcal{J}^{\mathrm{OI}}} w_{\alpha,(j,o,\tilde{j},i)} \dot{M}_{(j,o,\tilde{j},i)} \nonumber \\
&\quad + w_{\alpha,(j,o)}^{\mathrm{sink}} \dot{M}_{(j,o)}^{\mathrm{sink}} \quad \forall (j,o) \in \mathcal{J}^{\mathrm{O}}, \; \alpha \in \mathcal{C},
\end{align}
for each possible outlet port $(j,o)$ in the set $\mathcal{J}^{\mathrm{O}}$. For consistency, the mass fraction of each component $\alpha$ within all connecting mass streams and external sink flows originating from outlet $(j,o)$ is constrained to match the outlet composition:
\begin{align}
w_{\alpha,(j,o,\tilde{j},i)} &= w_{\alpha,(j,o)}^{\mathrm{out}} \quad \forall (j,o,\tilde{j},i) \in \mathcal{J}^{\mathrm{OI}}, \; \alpha \in \mathcal{C}, \\
w_{\alpha,(j,o)}^{\mathrm{sink}} &= w_{\alpha,(j,o)}^{\mathrm{out}} \quad \forall (j,o) \in \mathcal{J}^{\mathrm{O}}, \; \alpha \in \mathcal{C}.
\end{align}
This splitter formulation guarantees that mass fractions are consistently propagated across all material flow paths within and beyond system boundaries. The total mass flow exiting a process and distributed across all associated outlet ports is calculated \emph{via}:
\begin{align}
\dot{M}_{j}^{\mathrm{out,tot}} &= \sum_{o:(j,o) \in \mathcal{J}^{\mathrm{O}}} \dot{M}_{(j,o)}^{\mathrm{out}} \quad \forall j \in \mathcal{J}.
\end{align}
The component-wise distribution of the total outlet mass flow is described by:
\begin{align}
w_{\alpha,j}^{\mathrm{out,tot}} \dot{M}_{j}^{\mathrm{out,tot}} &= \sum_{o:(j,o) \in \mathcal{J}^{\mathrm{O}}} w_{\alpha,(j,o)}^{\mathrm{out}} \dot{M}_{(j,o)}^{\mathrm{out}} \quad \forall j \in \mathcal{J}, \; \alpha \in \mathcal{C}.
\end{align}
To ensure conservation of mass, the total mass flow exiting each process must equal the total mass flow entering that process:
\begin{align}
\dot{M}_{j}^{\mathrm{out,tot}} = \dot{M}_{j}^{\mathrm{in,tot}} \quad \forall j \in \mathcal{J}. 
\end{align}

Due to the occurrence of chemical conversions within individual processes, the mass composition of process streams is altered through the consumption or generation of chemical species. For those processes not represented \emph{via} ANNs, a simplified stoichiometry-based model is employed to approximate the relationship between reactants and products. Specifically, the component-wise total outlet mass flow is computed according to:
\begin{align}
w_{\alpha,j}^{\mathrm{out,tot}} \dot{M}_{j}^{\mathrm{out,tot}} = \, &w_{\alpha,j}^{\mathrm{in,tot}} \dot{M}_{j}^{\mathrm{in,tot}} + \nu_{\alpha,j} \Gamma_{j} \nonumber \\ 
&\forall j \in \mathcal{J} \setminus \mathcal{J}^{\mathrm{ANN}}, \; \alpha \in \mathcal{C}, \label{EQ: 13}
\end{align}
where $\nu_{\alpha,j}$ is the stoichiometric coefficient accounting for the net production ($\nu_{\alpha,j}  > 0$ for products) or consumption ($\nu_{\alpha,j} < 0$ for reactants) of component $\alpha$ in process $j$, and $\Gamma_{j}$ denotes the process capacity, \emph{i.e.}\@, the scale of process $j$. Thus, the reactant conversion or product generation is a linear function of the process capacity $\Gamma_{j}$. The parameter $\nu_{\alpha,j}$ and the operating conditions remain constant throughout, hence we refer to these processes as \emph{linear processes}. Based on the production rate of a designated key component $(\alpha,j)$, the process capacity $\Gamma_{j}$ associated with the process $j$ is calculated as follows:
\begin{align}
\Gamma_{j} = w_{(\alpha,j)}^{\mathrm{out,tot}} \dot{M}_{j}^{\mathrm{out,tot}} \quad \forall (\alpha,j) \in \mathcal{C}^{\mathrm{K}},  
\end{align}
where $\mathcal{C}^{\mathrm{K}}$ denotes the set of key components. For processes represented \emph{via} ANNs (collected in the set $\mathcal{J}^{\mathrm{ANN}}$), the outlet mass fractions are not computed from stoichiometric relations but are instead determined by the output of the corresponding trained machine learning models, which predict nonlinear relationships based on the given input variables. A detailed description of the ANN-based process modeling is provided in \autoref{Section: Process representation via neural networks}.

To ensure the physical consistency of mixture composition across all process streams, equality constraints are imposed on the summation of component mass fractions in each relevant flow. These constraints are expressed as:
\begin{align}
\sum_{\alpha \in \mathcal{C}} w_{\alpha,(j,i)}^{\mathrm{in}} &= y_{j} \quad \forall (j,i) \in \mathcal{J}^{\mathrm{I}}, \\
\sum_{\alpha \in \mathcal{C}} w_{\alpha,(j,o)}^{\mathrm{out}} &= y_{j} \quad \forall (j,o) \in \mathcal{J}^{\mathrm{O}}, \\
\sum_{\alpha \in \mathcal{C}} w_{\alpha,j}^{\mathrm{in,tot}} &= y_{j} \quad \forall j \in \mathcal{J}, \\
\sum_{\alpha \in \mathcal{C}} w_{\alpha,j}^{\mathrm{out,tot}} &= y_{j} \quad \forall j \in \mathcal{J}, \\
\sum_{\alpha \in \mathcal{C}} w_{\alpha,(j,o,\tilde{j},i)} &= y_{j} \quad \forall (j,o,\tilde{j},i) \in \mathcal{J}^{\mathrm{OI}}. 
\end{align}
The summation over all components $\alpha \in \mathcal{C}$ ensures that the total mass fractions in any given mass flow sum up to unity whenever the associated process is operational. The binary decision variable $y_{j} \in \{0,1\}$ encodes the status of process $j$, such that:
\begin{align*}
y_{j} =
\begin{cases}
1, & \text{if process $j$ is installed}, \\
0, & \text{otherwise},
\end{cases}
\quad &\forall j \in \mathcal{J}.
\end{align*}

\subsection{Electricity}
In addition to mass flows, processes within the superstructure are also characterized by the exchange of energy in the form of work (\emph{i.e.}\@, electricity) and heat, as illustrated in \autoref{Fig.: General process modeling.}. For all processes $j \in \mathcal{J} \setminus \mathcal{J}^{\mathrm{ANN}}$, which are not modeled \emph{via} ANNs, the work demand or generation rate $\dot{W}_j$ is formulated as a linear function of the process capacity $\Gamma_j$
\begin{align}
\dot{W}_j = e_j \Gamma_j \quad \forall j \in \mathcal{J} \setminus \mathcal{J}^{\mathrm{ANN}}, 
\end{align}
where $e_j$ denotes the specific work or electricity consumption/generation coefficient associated with process $j$. A negative value of $e_j$ implies net electrical energy consumption, whereas a positive value indicates net generation. For processes represented by ANNs (\emph{i.e.}\@, $j \in \mathcal{J}^{\mathrm{ANN}}$), the calculation of $\dot{W}_j$ is addressed in detail in \autoref{Section: Process representation via neural networks}. \autoref{EQ: 21} ensures that the sum of all generated and consumed work flows as well as the work that enters ($\dot{W}^{\mathrm{src}}$) the system from outside or exits ($\dot{W}^{\mathrm{sink}}$) the system \emph{via} the system boundaries fulfill the following condition:
\begin{align}
\sum_{j \in \mathcal{J}} \dot{W}_j + \dot{W}^{\mathrm{src}} - \dot{W}^{\mathrm{sink}} = 0. \label{EQ: 21}
\end{align}

\subsection{Heat integration} \label{Subsubsection: Heat integration}
Thermal integration constitutes a fundamental aspect of energy optimization in chemical process systems, as it enables the efficient redistribution of thermal energy between process streams. By facilitating internal system heat recovery, the dependence on external heating and cooling utilities is substantially reduced, thereby decreasing the overall energy consumption, operating expenditures, and associated greenhouse gas emissions. This enhances both the thermodynamic efficiency and environmental sustainability of the network. The present heat integration framework builds upon the work by Ganzer and Mac Dowell \cite{Ganzer.2020}, and is extended herein to accommodate the simultaneous representation of multiple thermal sources and sinks within a single process. This generalization allows for a more accurate and flexible depiction of complex heat exchange scenarios.

All processes possessing the potential for thermal integration are aggregated into the set $\mathcal{J}^{\mathrm{heat}}$. Furthermore, the set $\mathcal{J}^{\mathrm{heat,s}}$ encompasses all distinct heat source and sink elements within these processes that are eligible for participation in inter-process heat exchange. For processes not represented by ANNs, the thermal duty $\dot{Q}_{(j,h)}$ associated with each heat source or sink $(j,h)$ is computed as a function of the process capacity $\Gamma_j$ according to the relation:
\begin{align}
\dot{Q}_{(j,h)} = q_{(j,h)} \Gamma_j \quad \forall (j,h) \in \mathcal{J}^{\mathrm{heat,s}} \setminus \mathcal{J}^{\mathrm{ANN}},
\end{align}
where $q_{(j,h)}$ denotes the specific heat load parameter for the respective heat source or sink. A positive value of $q_{(j,h)}$ indicates heat release (source), while a negative value corresponds to heat demand (sink). Thermal energy balances are established for all process-internal heat sources and sinks to ensure conservation of heat across inter-process exchanges. For each heat source $(j,h)$, the balance equation is expressed as:
\begin{align}
\sum_{(\tilde{j}, \tilde{h}) \in \mathcal{J}^{\mathrm{heat,cold}}} &\dot{Q}_{(j,h),(\tilde{j},\tilde{h})}^{\mathrm{flow}} + \dot{Q}_{(j,h)}^{\mathrm{sink}} = \dot{Q}_{(j,h)} \nonumber 
\\ &\forall (j,h) \in \mathcal{J}^{\mathrm{heat,hot}} \setminus \{\mathrm{(Gasification},2)\}, \label{EQ: 23}
\end{align}
where $\dot{Q}_{(j,h),(\tilde{j},\tilde{h})}^{\mathrm{flow}}$ denotes the thermal energy transferred from heat source $(j,h)$ to heat sink $(\tilde{j},\tilde{h})$, and $\dot{Q}_{(j,h)}^{\mathrm{sink}}$ represents the proportion of thermal energy from source $(j,h)$ that exits the system boundaries. Similarly, for each heat sink, the balance is given by:
\begin{align}
\sum_{(\tilde{j}, \tilde{h}) \in \mathcal{J}^{\mathrm{heat,hot}}} &\dot{Q}_{(\tilde{j},\tilde{h}),(j,h)}^{\mathrm{flow}} + \dot{Q}_{(j,h)}^{\mathrm{src}} = -\dot{Q}_{(j,h)} \nonumber 
\\ &\forall (j,h) \in \mathcal{J}^{\mathrm{heat,cold}} \setminus \{\mathrm{(Gasification},2)\}. \label{EQ: 24}
\end{align}
where $\dot{Q}_{(j,h)}^{\mathrm{src}}$ accounts for the external thermal energy input required to satisfy the demand of sink $(j,h)$. The sets $\mathcal{J}^{\mathrm{heat,hot}}$ and $\mathcal{J}^{\mathrm{heat,cold}}$ contain all thermally active source and sink elements, respectively, within the superstructure network. A complete list of all processes participating in heat integration and all associated heat sources and sinks can be found in the SI (Subsection A.2.\@). Thermal energy exchanges between processes are constrained to be non-negative and are subject to upper bounds formulated using big-$M$ constraints. Specifically, the inter-process heat transfer $\dot{Q}_{(j,h),(\tilde{j},\tilde{h})}^{\mathrm{flow}}$ from a heat source $(j,h)$ to a heat sink $(\tilde{j},\tilde{h})$ must satisfy:
\begin{align}
\dot{Q}_{(j,h),(\tilde{j},\tilde{h})}^{\mathrm{flow}} \leq \, &z_{(j,h),(\tilde{j},\tilde{h})} \, M \nonumber \\ 
&\forall (j,h) \in \mathcal{J}^{\mathrm{heat,hot}}, \; (\tilde{j},\tilde{h}) \in \mathcal{J}^{\mathrm{heat,cold}},
\end{align}
where $M$ is a sufficiently large constant and $z_{(j,h),(\tilde{j},\tilde{h})} \in \{0,1\}$ is a binary decision variable that indicates whether a heat exchange link is active (\emph{i.e.}\@, equals 1 if heat is transferred, and 0 otherwise). Heat exchange is permitted only if both the supplying and receiving processes are operational. This is enforced by the following coupling constraints:
\begin{align}
z_{(j,h),(\tilde{j},\tilde{h})} \leq y_j \quad \forall (j,h) \in \mathcal{J}^{\mathrm{heat,hot}}, \\
z_{(j,h),(\tilde{j},\tilde{h})} \leq y_{\tilde{j}} \quad \forall (\tilde{j},\tilde{h}) \in \mathcal{J}^{\mathrm{heat,cold}}.
\end{align}
Furthermore, to maintain the thermodynamic feasibility of heat exchange, a minimum temperature driving force is imposed. Specifically, heat can only be transferred from a source to a sink if the source temperature exceeds the sink temperature by at least a specified threshold $\Delta T^{\mathrm{min}}$. This condition is represented as:
\begin{align}
T_{(\tilde{j},\tilde{h})} z_{(j,h),(\tilde{j},\tilde{h})} + &\Delta T^{\mathrm{min}} z_{(j,h),(\tilde{j},\tilde{h})} \leq T_{(j,h)} z_{(j,h),(\tilde{j},\tilde{h})} \nonumber \\ 
\forall &(j,h) \in \mathcal{J}^{\mathrm{heat,hot}}, \; (\tilde{j},\tilde{h}) \in \mathcal{J}^{\mathrm{heat,cold}}.
\end{align}
Here, $T_{(j,h)}$ is the temperature level associated with the heat port $(j,h)$, which is either a constant parameter (Table A.8.\@ in SI) or is determined as a decision variable of the optimization problem (more details are given in \autoref{Subsection: Superstruture optimization with embedded artificial neural networks}).

The equality constraints specified in \autoref{EQ: 23} and \autoref{EQ: 24} are not applicable to the thermal interaction unit $(j,h) = \{(\mathrm{Gasification},2)\}$, as this entity can function either as a heat source or a heat sink depending on the operating mode of the gasification step within the optimal system configuration. To accurately capture this dual-functionality within the superstructure optimization model, a dedicated set of constraints is introduced. Specifically, heat balance equations for the gasification process are reformulated as:
\begin{align}
&\sum_{(\tilde{j}, \tilde{h}) \in \mathcal{J}^{\mathrm{heat,cold}}} \dot{Q}_{(\mathrm{Gasification},2),(\tilde{j},\tilde{h})}^{\mathrm{flow}} + \dot{Q}_{(\mathrm{Gasification},2)}^{\mathrm{sink}} = \nonumber \\
&\qquad z^{\mathrm{src,sink}}_{(\mathrm{Gasification},2)} \dot{Q}_{(\mathrm{Gasification},2)}, 
\end{align}
\begin{align}
&\sum_{(\tilde{j}, \tilde{h}) \in \mathcal{J}^{\mathrm{heat,hot}}} \dot{Q}_{(\tilde{j},\tilde{h}),(\mathrm{Gasification},2)}^{\mathrm{flow}} + \dot{Q}_{(\mathrm{Gasification},2)}^{\mathrm{src}} = \nonumber \\ 
&\qquad \left(1 - z^{\mathrm{src,sink}}_{(\mathrm{Gasification},2)}\right) \left(-\dot{Q}_{(\mathrm{Gasification},2)}\right),
\end{align}
where the binary variable $z^{\mathrm{src,sink}}_{(\mathrm{Gasification},2)}$ is introduced to identify the operative mode of the gasification step: it assumes a value of 1 if $(j,h) = \{(\mathrm{Gasification},2)\}$ functions as a heat source, and 0 if it acts as a heat sink. To ensure consistency in this binary characterization, the heat duty of the gasification step is further bounded as
\begin{align}
&\dot{Q}_{(\mathrm{Gasification},2)} \leq z^{\mathrm{src,sink}}_{(\mathrm{Gasification},2)} M, \\
&\dot{Q}_{(\mathrm{Gasification},2)} \geq -\left(1 - z^{\mathrm{src,sink}}_{(\mathrm{Gasification},2)}\right) M.
\end{align}
Moreover, to avoid non-physical self-integration, the following constraint prohibits any thermal exchange from the gasification step to itself
\begin{align}
&z_{(\mathrm{Gasification},2),(\mathrm{Gasification},2)} = 0.
\end{align}
A comprehensive explanation of this bidirectional thermal port within the gasification process is provided in \autoref{Paragraph: Biomass gasification}.

\subsection{Economic constraints} \label{Subsubsection: Economic constraints}
The overarching objective of the optimization framework is the minimization of the total annualized cost of the network, formally expressed as
\begin{align}
K^{\mathrm{tot}} = CR \sum_{j \in \mathcal{J}} K^{\mathrm{cap}}_j + \sum_{\alpha \in \mathcal{C}^{\mathrm{src}}} K^{\mathrm{op,comp}}_{\alpha} + K^{\mathrm{op,el}} + K^{\mathrm{op,heat}},
\end{align}
where $K^{\mathrm{cap}}_j$ denotes the CAPEX associated with the installation of process $j$, $K^{\mathrm{op,comp}}_{\alpha}$ represents the operating cost linked to the procurement of compound $\alpha$, and $ K^{\mathrm{op,el}}$ and $K^{\mathrm{op,heat}}$ correspond to the annual electricity and heat-related operating costs, respectively. The term $CR$ is the capital recovery factor, which annualizes the investment costs over the project lifetime:
\begin{align}
CR = \frac{r (r + 1)^\theta}{(r + 1)^\theta - 1},
\end{align}
where a project lifetime of $\theta = 20$ years \cite{GonzalezGaray.2022} and an interest rate of $r = \SI{7}{\%}$ \cite{Svitnic.2022} are assumed. 

The CAPEX for processes that are not represented \emph{via} ANNs, with the exception of the acid gas removal (AGR) process, is calculated based on the process capacity $\Gamma_j$ \emph{via}:
\begin{align}
K^{\mathrm{cap}}_j = \beta_j \Gamma_j \quad \forall j \in \mathcal{J} \setminus \mathcal{J}^{\mathrm{ANN}} \setminus{\{\mathrm{AGR}\}}, \label{EQ: 36} 
\end{align}
where $\beta_j$ is the corresponding specific CAPEX of process $j$ in \$/capacity. For ANN-represented processes, where the process capacity $\Gamma_j$ is undefined, and for processes where there is a lack of cost data based on the process capacity, CAPEX is likewise estimated using the total inlet mass flow rate $\dot{M}^{\mathrm{in,tot}}$. This total inlet mass flow rate is scaled by process-specific investment cost factors sourced from relevant literature \cite{Rosenfeld.2020, GonzalezGaray.2022, Onel.2015}. The specific constraints for the CAPEX calculation based on the total inlet mass flow rate are listed in Subsection B.1.\@ of the SI.

OPEX associated with raw material procurement for each source component $\alpha \in \mathcal{C}^{\mathrm{src}}$ are computed based on the total mass flow rates of these components entering the designated process inlet ports. The corresponding cost expression is formulated as
\begin{align}
K^{\mathrm{op,comp}}_{\alpha} = \tau \gamma^{\mathrm{src}}_{\alpha} \sum_{(j,i) \in \mathcal{J}^{\mathrm{I}}} \dot{M}^{\mathrm{src}}_{\alpha,(j,i)} \quad \forall \alpha \in \mathcal{C}^{\mathrm{src}},
\end{align}
where $\tau = \SI{8760}{h}$ denotes the annual plant operating time (assuming continuous year-round operation) and $\gamma^{\mathrm{src}}_{\alpha}$ is the specific cost of raw material $\alpha$ in \$/kg. In a similar manner, the annual OPEX associated with external electricity and thermal energy supply are given by
\begin{align}
K^{\mathrm{op,el}} &= \tau \gamma^{\mathrm{el}} \dot{W}^{\mathrm{src}}, \\
K^{\mathrm{op,heat}} &= \tau \gamma^{\mathrm{heat}} \sum_{(j,h) \in \mathcal{J}^{\mathrm{heat,cold}}} \dot{Q}^{\mathrm{src}},
\end{align}
where $\gamma^{\mathrm{el}}$ and $\gamma^{\mathrm{heat}}$ denote the specific electricity and heat costs, respectively (in \$/kWh). Subsection A.3.\@ of the SI provides a comprehensive summary of all economic parameters.

\subsection{\ce{CO2} balancing} \label{Subsubsection: CO2 balancing}
In addition to the economic expenditures associated with the production of the kerosene fraction in the FT synthesis process, it is imperative to account for the total GHG emissions, quantified as \ce{CO2} equivalents (\ce{CO2}-eq), arising from the entire production pathway. The minimization of these \ce{CO2} emissions inherently conflicts with cost reduction, thereby introducing a multi-objective optimization problem involving a trade-off between economic and environmental performance metrics. The total \ce{CO2} emissions, denoted as $\dot{M}^{\mathrm{CO_2,tot}}$, are composed of several contributing factors
\begin{align}
\dot{M}^{\mathrm{CO_2,tot}} = &\sum_{\alpha \in \mathcal{C}^{\mathrm{src}}} \dot{M}^{\mathrm{CO_2,src}}_{\alpha} + \dot{M}^{\mathrm{CO_2,sink}} + \tau \lambda^{\mathrm{el}} \dot{W}^{\mathrm{src}} \nonumber \\
&+ \tau \lambda^{\mathrm{heat}} \sum_{(j,h) \in \mathcal{J}^{\mathrm{heat,cold}}} \dot{Q}^{\mathrm{src}}_{(j,h)}.
\end{align}
The parameters $\lambda^{\mathrm{el}}$ and $\lambda^{\mathrm{heat}}$ represent the specific \ce{CO2}-eq emission intensities associated with electricity and heat provision, respectively, measured in \si{\text{kg}\textsubscript{\ce{CO2}\text{-}eq}/\text{kWh}}. Detailed numerical values for these emission factors are provided in Subsection A.3.\@ of the SI.

The emissions attributed to raw material inputs are determined by:
\begin{align}
\dot{M}^{\mathrm{CO_2,src}}_{\alpha} = \tau \lambda^{\mathrm{comp}}_{\alpha} \sum_{(j,i) \in \mathcal{J}^{\mathrm{I}}} \dot{M}^{\mathrm{src}}_{\alpha,(j,i)} \quad \forall \alpha \in \mathcal{C}^{\mathrm{src}}, \label{EQ: 42}
\end{align}
where $\lambda^{\mathrm{comp}}_{\alpha}$ denotes the specific \ce{CO2}-eq emission factor associated with the supply chain and production of raw material $\alpha$. 

The term $\dot{M}^{\mathrm{CO_2,sink}}$ captures the net mass flow rate of \ce{CO2} that leaves the system boundary without being sequestered (sequestered carbon is stored permanently and does not leave the system boundaries) or utilized, and is computed as follows:
\begin{align}
\dot{M}^{\mathrm{CO_2,sink}} = &\tau \left( \sum_{(j,o) \in \mathcal{J}^{\mathrm{O}} \setminus \{\mathrm{CS,1}\}} w^{\mathrm{sink}}_{\mathrm{CO_2},(j,o)} \dot{M}^{\mathrm{sink}}_{(j,o)} \right. \nonumber \\
& + \left. \SI{1.571}{} \times \left( \sum_{(j,o) \in \mathcal{J}^{\mathrm{O}} \setminus\{\mathrm{FT,2}\}} w^{\mathrm{sink}}_{\mathrm{CO},(j,o)} \dot{M}^{\mathrm{sink}}_{(j,o)} \right) \right. \nonumber \\
& + \left. \SI{2.743} \times \left( \sum_{(j,o) \in \mathcal{J}^{\mathrm{O}} \setminus\{\mathrm{FT,1}\}} w^{\mathrm{sink}}_{\mathrm{CH_4},(j,o)} \dot{M}^{\mathrm{sink}}_{(j,o)} \right) \right).
\end{align}
where the coefficients \SI{1.571}{} and \SI{2.743}{} represent the molecular weight ratios for the complete conversion of \ce{CO} and \ce{CH4}, respectively, to \ce{CO2}, under the assumption that these species are fully oxidized upon release to the environment. The consideration of \ce{CO2} emissions due to the \ce{C}-contents in the product streams ($\mathrm{(FT,1)}$ and $\mathrm{(FT,2)}$) is explained later in \autoref{Subsubsection: Post optimization calculations}.

\subsection{Process-specific constraints}
Process-specific constraints are superimposed onto the general framework of the MIQCP formulation to enforce component-wise mass balances and ensure physicochemical feasibility of individual processes. These constraints restrict the admissible material flows such that only designated chemical species are permitted to enter or exit specific process streams in accordance with the underlying reaction stoichiometry and process function. For instance, within all electrolysis technologies, only \ce{H2O} is allowed as an input to the first inlet port, while the output streams are strictly limited to pure \ce{H2} and pure \ce{O2} at designated outlet ports. This behavior is mathematically enforced \emph{via} the following constraints and binary variables:
\begin{align}
&w^{\mathrm{in}}_{\mathrm{H_2O},(j,1)} = y_j, \quad \forall j = \{\mathrm{AEC, \, PEMEC, \, SOEC} \}, \label{EQ: 43} \\
&w^{\mathrm{out}}_{\mathrm{H_2},(j,1)} = y_j, \quad \; \; \forall j = \{\mathrm{AEC, \, PEMEC, \, SOEC} \}, \label{EQ: 44} \\
&w^{\mathrm{out}}_{\mathrm{O_2},(j,2)} = y_j, \quad \; \; \forall j = \{\mathrm{AEC, \, PEMEC, \, SOEC} \}. \label{EQ: 45}
\end{align}
These equality constraints ensure that the component-specific flow indicators are activated exclusively when the associated electrolysis process is selected for installation. Here, the abbreviations AEC, PEMEC, and SOEC stand for the three electrolyzer types considered in this study: alkaline electrolysis cells (AEC), proton exchange membrane electrolysis cells (PEMEC), and solid oxide electrolysis cells (SOEC). A detailed description of these processes is given later in \autoref{Paragraph: Water electrolysis}. Similar constraints to \autoref{EQ: 43}-\autoref{EQ: 45} are used to model the remaining superstructure optimization processes, with a full list of equations provided in Subsection B.1.\@ of the SI to this article.

To steer the optimization toward a desired production output, a fixed mass flow rate target is imposed on the primary outlet stream of the FT synthesis process. Specifically, the following constraint is introduced to enforce a fixed production rate:
\begin{align}
\dot{M}^{\mathrm{out}}_{\mathrm{(FT,1)}} = \SI{3000}{kg/h}.
\end{align}
In addition to meeting this quantitative production target, qualitative specifications pertaining to product composition are enforced to favor the generation of the desired \ce{C8}–\ce{C16} kerosene hydrocarbon fraction. This compositional tuning is realized \emph{via} inequality constraints that suppress the relative formation of undesired byproducts (namely, lighter gasoline-range hydrocarbons and heavier diesel-range compounds) by requiring their combined mass fractions (along with waste water) not to exceed that of the kerosene fraction in the FT product stream
\begin{align}
\sum_{\alpha \in \mathcal{C}^{\mathrm{gasoline}}} w^{\mathrm{out}}_{\alpha,\mathrm{(FT,1)}} + w^{\mathrm{out}}_{\mathrm{H_2O_{waste},(FT,1)}} &\leq \sum_{\alpha \in \mathcal{C}^{\mathrm{kerosene}}} w^{\mathrm{out}}_{\alpha,\mathrm{(FT,1)}}, \\
\sum_{\alpha \in \mathcal{C}^{\mathrm{diesel}}} w^{\mathrm{out}}_{\alpha,\mathrm{(FT,1)}} + w^{\mathrm{out}}_{\mathrm{H_2O_{waste},(FT,1)}} &\leq \sum_{\alpha \in \mathcal{C}^{\mathrm{kerosene}}} w^{\mathrm{out}}_{\alpha,\mathrm{(FT,1)}}.
\end{align}
These constraints collectively steer the optimization toward maximizing the yield of the kerosene-range hydrocarbons, thereby enhancing the overall selectivity in favor of the target product fraction.

\subsection{Post-optimization calculations} \label{Subsubsection: Post optimization calculations}
To rigorously attribute both cost and \ce{CO2} emissions to the production of the \ce{C8}–\ce{C16} kerosene fraction generated \emph{via} the FT synthesis, a series of post-optimization calculations are employed. These computations utilize the optimal decision variables obtained from the solved MIQCP problem. The rationale for performing these evaluations \emph{a posteriori} lies in the fact that integrating them directly into the optimization model would substantially increase the degree of nonlinearity, thereby exacerbating the computational complexity.

The total annual \ce{CO2} emissions attributable to the production and combustion of the kerosene fraction are quantified \emph{via} the following relationship:
\begin{align}
\dot{M}^{\mathrm{CO_2,tot,kerosene}} = \frac{\delta^{\mathrm{kerosene}}}{\delta^{\mathrm{HC}}} \dot{M}^{\mathrm{CO_2,tot}} + \tau \lambda^{\mathrm{kerosene}} \dot{M}^{\mathrm{out}}_{\mathrm{(FT,1)}} \nonumber \\
\quad + \tau \sum_{\alpha \in \mathcal{C}^{\mathrm{kerosene}}} w^{\mathrm{sink}}_{\alpha,\mathrm{(FT,1)}} \times \left(\SI{1.571}{} \times w^{\mathrm{sink}}_{\mathrm{CO,(FT,2)}} \dot{M}^{\mathrm{sink}}_{\mathrm{(FT,2)}}\right), \label{EQ: 49}
\end{align}
where emissions are allocated based on the energy-equivalent economic value of the kerosene fraction. This value is represented by the ratio of the specific energy content of the kerosene fraction, $\delta^{\mathrm{kerosene}}$, to the total specific energy content of all hydrocarbons produced by the FT process, $\delta^{\mathrm{HC}}$. Incorporating constraint \autoref{EQ: 49} into the original optimization problem would cause this quotient in particular to significantly increase the degree of nonlinearity, yielding an MINLP instead of MIQCP and rendering optimization more challenging. The specific energy values are computed as weighted sums based on the mass fractions and specific lower heating values ($\delta_{\alpha}$) of the individual hydrocarbon components
\begin{align}
\delta^{\mathrm{kerosene}} &= \sum_{\alpha \in \mathcal{C}^{\mathrm{kerosene}}} w^{\mathrm{out}}_{\alpha,\mathrm{(FT,1)}} \delta_{\alpha}, \\
\delta^{\mathrm{HC}} &= \sum_{\alpha \in \mathcal{C}^{\mathrm{HC}}} w^{\mathrm{out}}_{\alpha,\mathrm{(FT,1)}} \delta_{\alpha}.
\end{align}
Combustion-related \ce{CO2} emissions are accounted for \emph{via} the term $\lambda^{\mathrm{kerosene}}$, which reflects the specific \ce{CO2} emissions per kilogram of kerosene combusted and is defined as:
\begin{align}
\lambda^{\mathrm{kerosene}} &= \sum_{\alpha \in \mathcal{C}^{\mathrm{kerosene}}} w^{\mathrm{out}}_{\alpha,\mathrm{(FT,1)}} \lambda^{\mathrm{comp}}_{\alpha},
\end{align}
assuming complete oxidation of the hydrocarbon constituents. Additionally, unconverted and purged \ce{CO} from the second outlet port of the FT reactor is proportionally assigned to the kerosene product stream under the conservative assumption of complete atmospheric oxidation to \ce{CO2}.

In parallel with the emission allocation, total annualized production costs are similarly attributed to the kerosene fraction based on the same energy-weighted economic allocation principle
\begin{align}
K^{\mathrm{tot,kerosene}} = \frac{\delta^{\mathrm{kerosene}}}{\delta^{\mathrm{HC}}} K^{\mathrm{tot}}.
\end{align}

The total annual mass of kerosene-range hydrocarbons produced is computed as:
\begin{align}
\dot{M}^{\mathrm{sink,kerosene}} = \tau \sum_{\alpha \in \mathcal{C}^{\mathrm{kerosene}}} w^{\mathrm{sink}}_{\alpha,\mathrm{(FT,1)}} \dot{M}^{\mathrm{out}}_{\mathrm{(FT,1)}}.
\end{align}
From these values, the specific \ce{CO2} emissions and specific production costs per unit mass of kerosene (in \si{kg_{\ce{CO2}}/kg_{kerosene}} and \si{\$/kg_{kerosene}}, respectively) are obtained through normalization by the annual kerosene output.

The economic viability and environmental effectiveness of alternative process configurations or technologies relative to a fossil-based benchmark can be quantitatively assessed through the metric of \ce{CO2} abatement cost, denoted as $k^{\mathrm{ab.}}_{\mathrm{CO_2}}$. This value is defined as the incremental cost per unit of \ce{CO2} avoided and is computed according to the following relationship \cite{Salem.2023}:
\begin{align}
k^{\mathrm{ab.}}_{\mathrm{CO_2}} \left(\frac{\$}{\mathrm{t_{CO_2}}}\right) = \frac{k_{\mathrm{green \; kerosene}} - k_{\mathrm{reference}} \left(\frac{\$}{\mathrm{t_{kerosene}}}\right)}{\mathrm{CO_2 \; avoidance} \left(\frac{\mathrm{t_{CO_2}}}{\mathrm{t_{kerosene}}} \right)}, \label{EQ: 59}
\end{align}
where $k$ represents the specific production cost (\si{\$/kg\textsubscript{kerosene}}) and "abatement" is abbreviated as "ab". This metric enables a direct comparison across different pathways and technologies, facilitating the identification of the most cost-effective options.

A list of all parameters can be found in Subsection A.3.\@ of the SI to this article.

\section{Data set for Fischer-Tropsch sustainable aviation fuel} \label{Section: Data set for Fischer-Tropsch sustainable aviation fuel}
\begin{figure*}[h]
\centering
  \includegraphics[trim=0.5cm 1.25cm 0.5cm 1.5cm, clip, width=1\columnwidth]{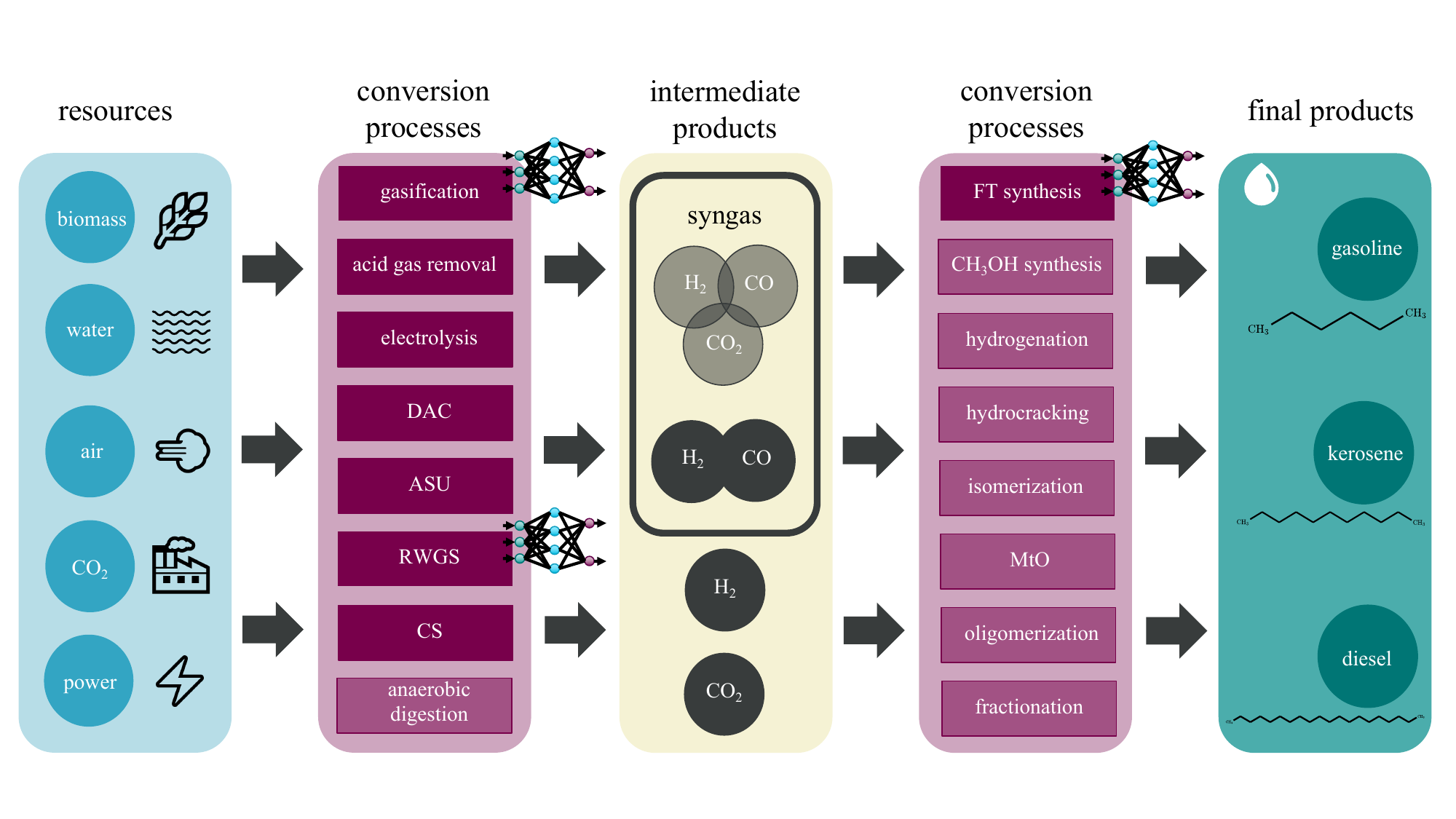}
\caption{Simplified representation of the superstructure network for the production of kerosene \emph{via} Fischer-Tropsch (FT) synthesis. Processes that are represented by an artificial neural network (ANN) are indicated. Components and processes not considered in this study are greyed out. (DAC: direct air capture; ASU: air separation unit; RWGS: reverse water-gas shift; CS: carbon sequestration; FT: Fischer-Tropsch; MtO: Methanol-to-Olefins).}
\label{Fig.: Superstructure network for the production of kerosene fraction as Fischer-Tropsch (FT) output}
\end{figure*}

Conventional aviation fuels for commercial passenger aircraft, such as Jet A-1, are complex multicomponent mixtures comprising predominantly n-alkanes, iso-alkanes, cycloalkanes, and aromatic hydrocarbons \cite{Holladay.2020}. In contrast, SAFs derived from renewable feedstocks \emph{via} FT synthesis exhibit a hydrocarbon distribution largely dominated by linear n-alkanes \cite{Bauen.2020}. Owing to their limited compositional diversity and associated physicochemical properties, these FT-derived SAFs typically require blending with conventional petroleum-based kerosene to meet the stringent performance and compatibility criteria for combustion in commercial aviation turbines. Given the methodological focus of the present study, the scope is restricted to the synthesis of n-alkanes in the carbon range \ce{C1}–\ce{C30} as the representative FT product spectrum. Alternative production pathways and post-synthesis upgrading processes, such as hydrocracking and isomerization, are not considered within this work. The superstructure network for n-alkane synthesis \emph{via} the FT route, encompassing raw material inputs, intermediate processing stages, and final product mixtures, is schematically illustrated in \autoref{Fig.: Superstructure network for the production of kerosene fraction as Fischer-Tropsch (FT) output}. A brief overview of the components and subprocesses is provided in the following sections.

\subsection{Components}
In the context of the superstructure optimization, components are defined as all chemical species that may be consumed, produced, or transformed by the various processes within the system. These include species that either enter the system boundaries as external resources (designated as source components) or exit the network (referred to as sink components). Representative feedstocks considered in this work encompass three lignocellulosic biomass types (miscanthus, wheat straw, and pine chips), as well as inorganic inputs such as air and water. Syngas, comprising predominantly \ce{CO} and \ce{H2}, is generated as an intermediate mixture for subsequent synthesis. The product spectrum includes straight-chain alkanes (\ce{C1} to \ce{C30}), which are modeled as final output components. To account for hydrocarbons beyond this range, a pseudo-component denoted as $\mathrm{C_{30+}}$ is introduced, aggregating all longer-chain n-alkanes. The system comprises a total of \SI{47}{} distinct components; a comprehensive enumeration is provided in Subsection A.2.\@ of the SI.

\subsection{Processes} \label{Subsection: Processes}
Processes convert one or more input components into a set of output components. Within the scope of this study, a total of eleven distinct processes are incorporated into the superstructure network. A key differentiation among these lies in the modeling approach: most processes are represented using simplified linear relationships, commonly referred to as short-cut models, which approximate conversion mechanisms and have been the standard in nearly all previous work. In contrast, three processes whose behavior changes significantly for varying operating conditions and inlet compositions are described by ANNs, allowing a more detailed representation beyond the limits of short-cut approximations. A detailed description of these processes is provided in the following, the surrogate training and the integration of the ANNs are discussed later in \autoref{Section: Process representation via neural networks}. Subsection A.2.\@ of the SI presents an overview and classification of all processes.

\subsubsection{Processes represented by artificial neural networks} \label{Subsubsection: Processes represented by artificial neural networks}
To achieve an optimal trade-off between the accurate representation of complex nonlinear process behavior through ANNs and the computational tractability of the overall superstructure optimization, only those processes which exhibit strong nonlinearities and where changes in operating conditions change the material flows in the superstructure are represented as ANNs. Aspen Plus\textsuperscript{\textregistered} \cite{AspenPlu.24.08.2023} simulations serve as the basis for generating the datasets used to train the surrogate models. Specifically, the biomass gasification, reverse water-gas shift (RWGS), and FT synthesis processes are represented \emph{via} rigorous flowsheet simulations to capture the dependence of their output compositions on varying operating conditions, such as temperature, pressure, and inlet compositions.

\paragraph*{Biomass gasification.} \label{Paragraph: Biomass gasification}
Gasification constitutes a complex, multi-step thermochemical conversion pathway wherein lignocellulosic biomass is transformed into a gaseous product mixture predominantly composed of \ce{H2}, \ce{CO}, \ce{CO2}, and water vapor (\ce{H2O}). Minor constituents, including nitrogen (\ce{N2}), ammonia (\ce{NH3}), hydrogen chloride (\ce{HCl}), hydrogen sulfide (\ce{H2S}), and condensable hydrocarbons such as tars, are also typically present as process-derived impurities \cite{Doherty.2013, Tanzil.2021}. The overall gasification mechanism encompasses several sequential and overlapping reactions, notably pyrolysis, partial oxidation, and hydrogenation. Pyrolytic decomposition generally occurs at temperatures below \SI{600}{\degreeCelsius}, while partial oxidation and associated hydrogenation reactions proceed at significantly elevated temperatures, ranging from \SI{700}{} to \SI{1600}{\degreeCelsius}. To avoid complete combustion of carbonaceous species to \ce{CO2}, the process operates at sub-stoichiometric oxygen levels. The selection of the oxidizing medium -- steam, air, pure \ce{O2}, or \ce{CO2} -- significantly alters the thermodynamic driving forces and dominant reaction pathways. For instance, steam-enhanced gasification promotes hydrogen formation \emph{via} the water-gas shift equilibrium, whereas oxygen-enriched environments favor syngas generation with increased lower heating values. Additionally, \ce{CO2}-assisted gasification can enhance CO production through the Boudouard reaction. The physicochemical characteristics of the biomass feedstock (\emph{e.g.}\@, miscanthus, wheat straw, or pine chips) directly affect the yield and speciation of by-products, particularly nitrogen- and sulfur-containing compounds, which necessitate downstream gas-cleaning measures prior to FT synthesis \cite{Yang.2018}.

\begin{figure*}[h]
\centering
  \includegraphics[trim=0.8cm 3cm 0.5cm 3.2cm, clip, width=1\columnwidth]{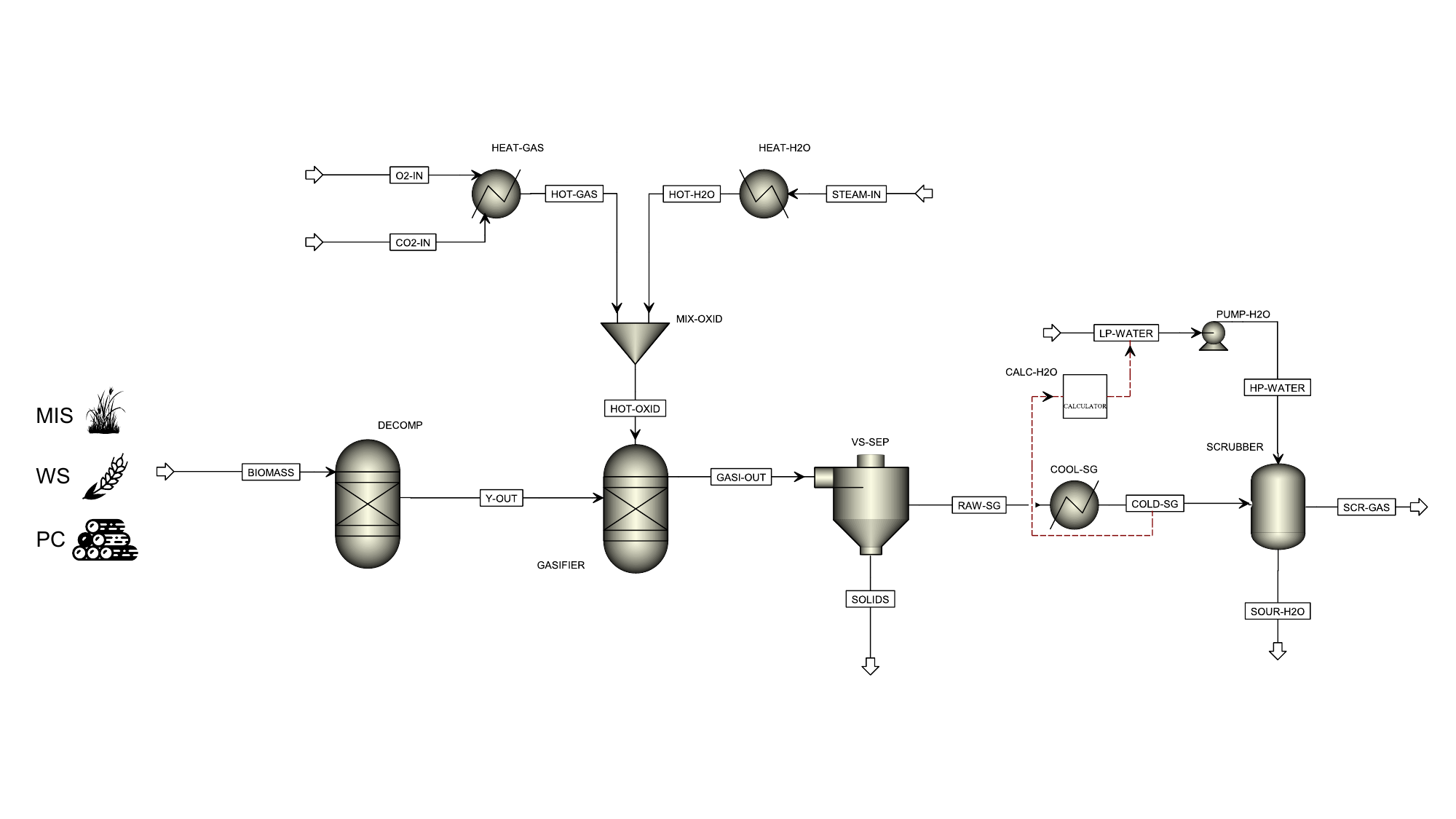}
\caption{Flowsheet of the gasification process, simulated in Aspen Plus\textsuperscript{\textregistered} \cite{AspenPlu.24.08.2023}, taking into account three types of biomass. The gasification process is divided into three principal stages: thermal decomposition and gasification of the biomass feedstock, pretreatment of the gasifying agents, and subsequent gas purification (removal of solid components and other impurities). (MIS: miscanthus; WS: wheat straw; PC: pine chips).}
\label{Fig.: Aspen Plus flowsheet simulation of gasification process.}
\end{figure*}

\autoref{Fig.: Aspen Plus flowsheet simulation of gasification process.} shows the Aspen Plus\textsuperscript{\textregistered} \cite{AspenPlu.24.08.2023} flowsheet of the biomass gasification process.
Peng-Robinson-Boston-Mathias (PR-BM) is chosen as the global property method for this model \cite{Peng.1976, boston1980proceedings}. Biomass and ash are treated as non-conventional solids based on the proximate, ultimate and sulphur analyses, while Aspen Plus\textsuperscript{\textregistered} \cite{AspenPlu.24.08.2023} proprietary methods DCOALIGT and HCOALGEN are used for density and enthalpy calculations \cite{KT.2023}. 

The gasification process is conceptually divided into three principal stages: (i) thermal decomposition and gasification of the biomass feedstock, (ii) pretreatment of the gasifying agents, and (iii) subsequent gas purification. In this study, three distinct lignocellulosic biomass types (miscanthus, wheat straw, and pine chips), characterized by varying proximate and ultimate compositions (detailed in Table C.1.\@ in the SI), are considered to capture feedstock heterogeneity. The initial decomposition phase involves pyrolyzing the biomass in an inert environment at \SI{450}{\degreeCelsius}, in the absence of gasification agents. This devolatilization step is modeled using the RYield reactor module in Aspen Plus\textsuperscript{\textregistered} \cite{AspenPlu.24.08.2023}, which disaggregates the biomass into volatile gaseous species and residual solid char as a function of its composition \cite{KT.2023, Yang.2018}.
%
%
Following the initial decomposition, the resultant biomass fragments are introduced into the gasification reactor alongside preheated oxidizing agents (namely steam, \ce{O2}, and/or \ce{CO2}) at an inlet temperature of \SI{450}{\degreeCelsius}. Gasification is modeled using a thermodynamic equilibrium-based approach \emph{via} an RGibbs reactor in Aspen Plus\textsuperscript{\textregistered} \cite{AspenPlu.24.08.2023}, operating within a temperature range of \SI{800}{} to \SI{1300}{\degreeCelsius}. The overall thermal nature of this stage, either exothermic or endothermic, is dependent upon both the reactor temperature and the molar quantities of oxidizing agents, thereby dictating whether auxiliary heat input is required. Post-gasification, residual solids are removed \emph{via} a SSplit module, configured with a split fraction of one for solid species to ensure complete separation. The remaining gaseous stream is then cooled to \SI{300}{\degreeCelsius} prior to undergoing a gas purification step, where a water scrubbing unit eliminates soluble contaminants, particularly ammonia (\ce{NH3}) \cite{Yang.2018}. Inputs for the flowsheet simulation and subsequent training of the gasification ANN with the associated bounds can be found in \autoref{Table: Inputs of the biomass gasification ANN with the corresponding upper and lower bounds}.
\begin{table}[h]
\small
\caption{Inputs of the artificial neural network (ANN) representing the biomass gasification process with the corresponding upper and lower bounds. (MIS: miscanthus; WS: wheat straw; PC: pine chips).}
\label{Table: Inputs of the biomass gasification ANN with the corresponding upper and lower bounds}
\begin{tabular*}{1\textwidth}{@{\extracolsep{\fill}}l r r} 
\hline
\multicolumn{1}{l}{ANN input} & \multicolumn{1}{c}{lower bound} & \multicolumn{1}{c}{upper bound} \\
\hline
biomass type & \multicolumn{2}{c}{$\{$MIS, WS, PC$\}$} \\
\hline
steam-to-biomass ratio (mass-based) & 0.01 & 1 \\
\ce{CO2}-to-biomass ratio (mass-based) & 0 & 1 \\
\ce{O2}-to-biomass ratio (mass-based) & 0.05 & 0.3 \\
gasifier temperature & \SI{800}{\degreeCelsius} & \SI{1300}{\degreeCelsius} \\
\hline
\end{tabular*}
\end{table}

To incorporate the distinct biomass feedstocks within the gasification ANN \emph{via} one-hot encoding, an auxiliary set of binary decision variables is introduced into the MIQCP formulation. These variables are defined as follows:
\begin{align*}
x_{\alpha} =
\begin{cases}
1, & \text{if biomass type $\alpha$ is selected}, \\
0, & \text{otherwise},
\end{cases}
\quad &\forall \alpha \in \mathcal{C}^{\mathrm{biomass}},
\end{align*}
where $\mathcal{C}^{\mathrm{biomass}}$ denotes the set of biomass types (\emph{e.g.}\@, miscanthus, wheat straw, and pine chips).

To align with the generalized process configuration depicted in \autoref{Fig.: General process modeling.}, the gasification process flowsheet is designed with five distinct inlet ports: one for the biomass feedstock, three for the oxidizing agents (steam, \ce{O2}, and \ce{CO2}), and one for the water supply to the downstream scrubbing unit. The system also incorporates three outlet ports: a solid-phase discharge stream, a waste water stream from the scrubbing unit (containing dissolved contaminants), and a gas-phase product stream representing the cleaned synthesis gas exiting the process. Thermal integration within the gasification subsystem is modeled through three heat exchange interfaces. The decomposition stage necessitates a thermal input at approximately \SI{450}{\degreeCelsius}, while thermal energy is recuperated during the post-gasification cooling phase at \SI{300}{\degreeCelsius}. Additionally, the gasification reactor itself may operate as either an exothermic or endothermic unit, functioning as a heat source or sink depending on the specific operating conditions and oxidant loading, as elaborated in \autoref{Subsubsection: Heat integration}.

Overall, the ability of the gasification process to operate under various conditions enables it to adapt to changing circumstances (\emph{e.g.}\@, different inlet composition, raw material costs) in the superstructure.

\paragraph*{Reverse water-gas shift (RWGS).}
\ce{CO2} captured from air must undergo chemical activation prior to its utilization in the synthesis of long-chain hydrocarbons \emph{via} FT synthesis. This activation is achieved through the RWGS process, wherein \ce{CO2} is reacted with \ce{H2} to form \ce{CO} and \ce{H2O}. The RWGS reaction is endothermic, as described by the following stoichiometry:
\begin{align*}
\ce{CO2} + \ce{H2} \rightleftharpoons \ce{CO} + \ce{H2O} \quad \Delta_{\mathrm{R}}\tilde{h}^{\theta} = \SI{41}{kJ/mol},
\end{align*}
and is typically favored at elevated temperatures in the range of \SI{800}{} to \SI{900}{\degreeCelsius}, where equilibrium conversion is thermodynamically enhanced \cite{Elsernagawy.2020, GonzalezGaray.2022}. The resulting syngas can then be supplied to the FT synthesis.

\begin{figure*}[h]
\centering
  \includegraphics[trim=3.7cm 10cm 1.2cm 10cm, clip, width=1\columnwidth]{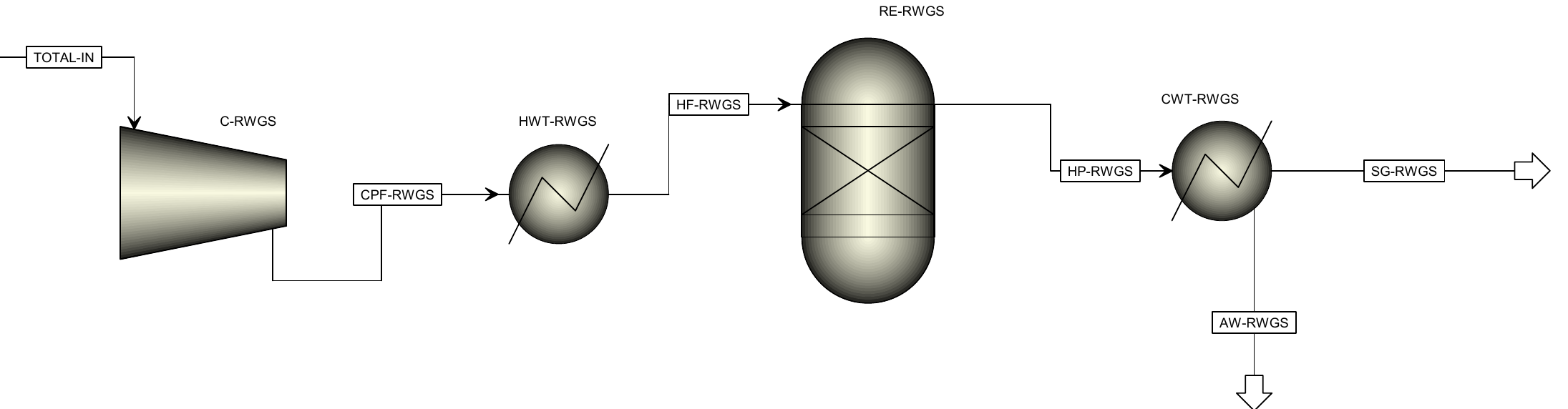}
\caption{Flowsheet of the reverse water-gas shift (RWGS) process, simulated in Aspen Plus\textsuperscript{\textregistered} \cite{AspenPlu.24.08.2023}. The RWGS process consists of three parts: pretreatment of the feed mixture, the RWGS equilibrium reaction, and cooling of the reaction mixture and associated water separation.}
\label{Fig.: Aspen Plus flowsheet simulation of RWGS process.}
\end{figure*}

The RWGS process flow diagram is presented in \autoref{Fig.: Aspen Plus flowsheet simulation of RWGS process.}.
Following Bube \emph{et al.}\@ \cite{Bube.2024}, the PR-BM equation of state is employed as the global property method to ensure accurate thermophysical property estimation across the relevant pressure and temperature regimes. A mixture of \ce{CO2} and \ce{H2} is first subjected to polytropic compression to attain the designated operating pressure of \SI{20}{bar}. The compressed gas mixture is subsequently heated to the target reaction temperature. The endothermic RWGS reaction is simulated within an RGibbs reactor unit under thermodynamic equilibrium conditions, with reaction temperatures ranging from \SI{850}{} to \SI{1000}{\degreeCelsius}. Continuous external heat input is required to sustain isothermal reactor operation. Nickel-based catalysts are typically employed in practical applications to enhance the reaction kinetics and facilitate favorable equilibrium conversion \cite{Bube.2024, Elsernagawy.2020}. Following the reaction, the effluent gas stream, comprising \ce{CO2}, \ce{H2}, \ce{CO}, and \ce{H2O}, is cooled to near-ambient conditions, allowing for the removal of condensed water \emph{via} phase separation. The inputs and the corresponding lower and upper bounds for the RWGS ANN are listed in \autoref{Table: Inputs of the RWGS ANN with the corresponding upper and lower bounds}.
\begin{table}[h]
\small
\caption{Inputs of the artificial neural network (ANN) representing the reverse water-gas shift (RWGS) process with the corresponding upper and lower bounds.}
\label{Table: Inputs of the RWGS ANN with the corresponding upper and lower bounds}
\begin{tabular*}{1\textwidth}{@{\extracolsep{\fill}}l r r} 
\hline
\multicolumn{1}{l}{ANN input} & \multicolumn{1}{c}{lower bound} & \multicolumn{1}{c}{upper bound} \\
\hline
reactor temperature & \SI{850}{\degreeCelsius} & \SI{1000}{\degreeCelsius} \\
\ce{H2} mass fraction in & 0.02 & 0.25 \\
\hline
\end{tabular*}
\end{table}

From a process integration standpoint, the RWGS process comprises a single inlet (reactant mixture), two outlet ports (synthesis gas and liquid water), one heat sink (representing the aggregate thermal demand for both preheating and isothermal reactor operation), and one heat source (associated with post-reaction cooling and water condensation).

\paragraph*{Fischer-Tropsch (FT) synthesis.}
FT synthesis encompasses a series of polymerization reactions whereby synthesis gas is catalytically converted predominantly into linear alkanes spanning a broad carbon number distribution, from light gaseous species such as methane (\ce{C1}) to high molecular weight waxes. This transformation can be represented by this generalized stoichiometric reaction:
\begin{align*}
\mathrm{n} \, \ce{CO} + \mathrm{(2n+1)} \, \ce{H2} \rightarrow \mathrm{C_nH_{2n+2}} + \mathrm{n} \, \ce{H2O},
\end{align*}
where $n$ denotes the carbon chain length \cite{Bube.2024, Do.2022}. In industrial settings, minor fractions of alkenes and oxygenated hydrocarbons are concurrently produced due to reaction pathways and limited catalyst selectivity. The product distribution is modulated by catalyst composition, primarily cobalt- or iron-based catalysts, and reaction conditions, including temperature, pressure, and synthesis gas feed composition \cite{Maitlis.2013}. Within the present investigation, the FT synthesis output constitutes the principal target product stream, with a design objective to maximize yield within the \ce{C8} to \ce{C16} carbon range, corresponding to optimal hydrocarbon chain lengths for SAF applications \cite{Holladay.2020}.

\begin{figure*}[h]
\centering
  \includegraphics[trim=2.5cm 7cm 1cm 7cm, clip, width=1\columnwidth]{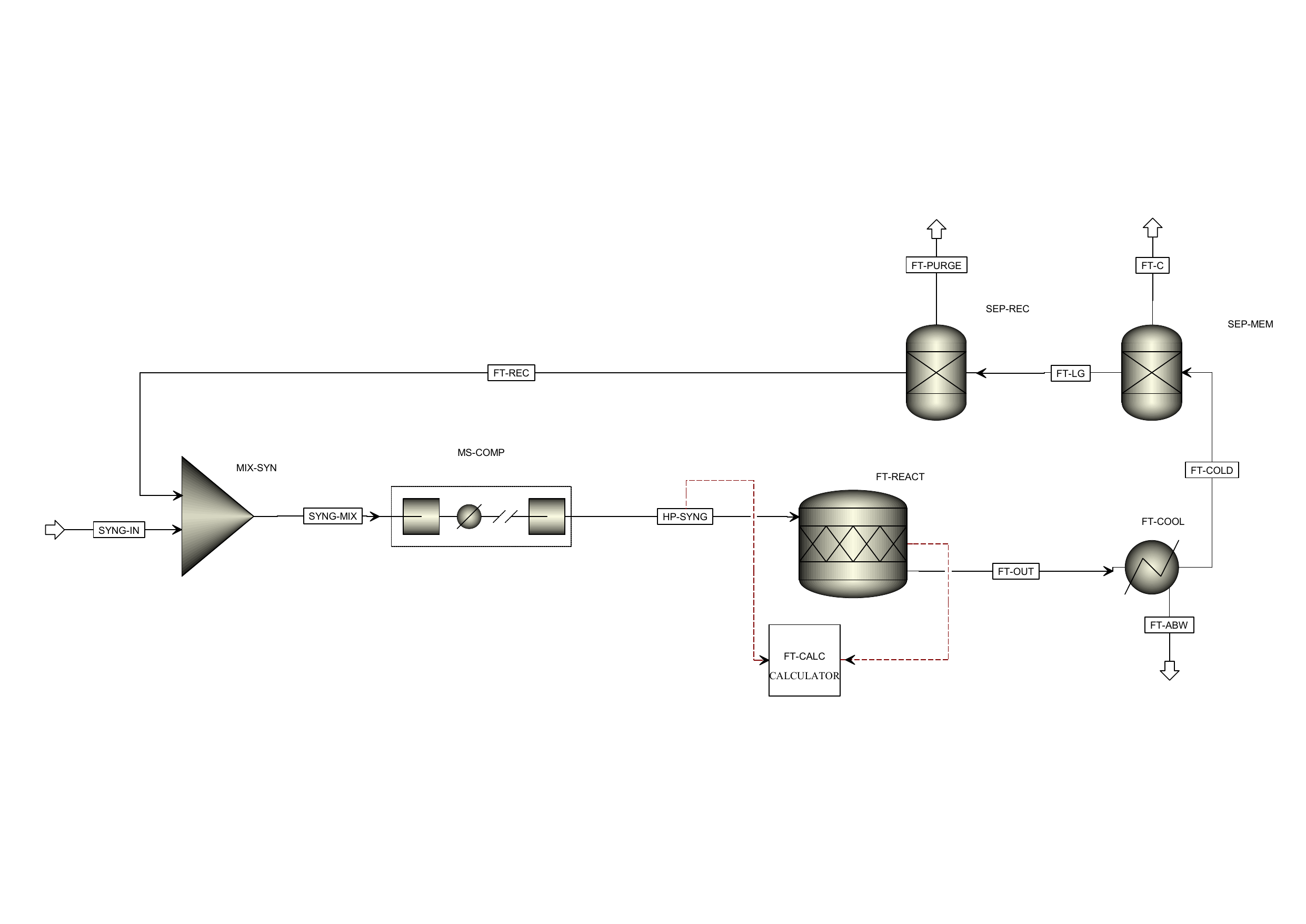}
\caption{Flowsheet of the Fischer-Tropsch (FT) process, simulated in Aspen Plus\textsuperscript{\textregistered} \cite{AspenPlu.24.08.2023}. The FT process consists of the following steps: pretreatment of the synthesis gas mixture, the exothermic FT reaction, product separation (water removal, separation of light gas components from long-chain hydrocarbons), and recycling of the reactant components \ce{CO} and \ce{H2}.}
\label{Fig.: Aspen Plus flowsheet simulation of FT process.}
\end{figure*}

The chain length distribution of the FT synthesis exhibits significant variability as a function of pressure, temperature, and the \ce{H2}/\ce{CO} feed ratio \cite{SunFT.2017, Yang.2017}. To capture this complex, nonlinear behavior within the context of our optimization framework, an ANN is developed and trained utilizing simulation data generated from an Aspen Plus\textsuperscript{\textregistered} \cite{AspenPlu.24.08.2023} model, as illustrated in \autoref{Fig.: Aspen Plus flowsheet simulation of FT process.}.
Following the other two steady-state flowsheet simulations (gasification and RWGS), PR-BM is chosen as the global property method \cite{Bube.2024}. 

Initially, high-purity synthesis gas is blended with recycled, unconverted \ce{CO} and \ce{H2} prior to undergoing compression. The compression stage is modeled as a multi-stage compression system incorporating interstage cooling, wherein the gas mixture is elevated to an operational pressure in the range of \SI{30}{} to \SI{55}{bar}, with the discharge temperature regulated to align with the desired reaction temperature, set between \SI{200}{} and \SI{300}{\degreeCelsius} \cite{Maitlis.2013}. Subsequently, the gas stream enters the FT synthesis stage, where we assume cobalt-catalyzed, low-temperature FT synthesis due to its favorable kinetics and selectivity for long-chain paraffinic hydrocarbons. The FT reactor is modeled as a stoichiometric reactor (RStoic) within Aspen Plus\textsuperscript{\textregistered} \cite{AspenPlu.24.08.2023}, encompassing a detailed reaction scheme that accounts for the formation of linear n-alkanes ranging from \ce{C1} to \ce{C30}. To approximate the contribution of higher molecular weight fractions, a lumped pseudo-component, denoted as $\mathrm{C_{30+}}$, is introduced. Minor by-products such as olefins and oxygenates are excluded from the model owing to their low yield under cobalt catalysis and limited relevance to SAF production \cite{Hamelinck.2004, Klerk.2011}. The hydrocarbon product distribution is computed based on an empirically derived chain growth probability parameter, utilizing the Anderson–Schulz–Flory (ASF) distribution model. This distribution is dependent on key operating variables, including reaction pressure, temperature, and syngas composition \cite{Hamelinck.2004, Klerk.2011}. Details of the custom Excel-calculator block routine integrated within our Aspen Plus\textsuperscript{\textregistered} \cite{AspenPlu.24.08.2023} simulation are provided in Subsection C.2.\@ in the SI. Post-reaction, the effluent is subjected to cooling to facilitate condensation and removal of the majority of water generated during synthesis. The product stream is subsequently separated from unreacted syngas using simplified split fraction models. A portion of the unconverted \ce{CO}/\ce{H2} stream is purged to prevent buildup within the system, while the remainder is recycled and remixed with fresh feed. Approximately \SI{98}{\%} of unreacted \ce{H2} is recovered \emph{via} recycling. The \ce{CO} recycle rate is adjusted \emph{via} a design specification by adapting the purge ratio to maintain the \ce{H2}/\ce{CO} molar ratio at the FT reactor inlet within the valid range of the empirical FT reactor model. \autoref{Table: Inputs of the FT ANN with the corresponding upper and lower bounds}
\begin{table}[h]
\small
\caption{Inputs of the artificial neural network (ANN) representing the Fischer-Tropsch (FT) process with the corresponding upper and lower bounds.}
\label{Table: Inputs of the FT ANN with the corresponding upper and lower bounds}
\begin{tabular*}{1\textwidth}{@{\extracolsep{\fill}}l r r}
\hline
\multicolumn{1}{l}{ANN input} & \multicolumn{1}{c}{lower bound} & \multicolumn{1}{c}{upper bound} \\
\hline
reactor temperature & \SI{200}{\degreeCelsius} & \SI{300}{\degreeCelsius} \\
reactor pressure & \SI{30}{bar} & \SI{55}{bar} \\
\ce{H2} mass fraction in & 0.115 & 0.15 \\
\hline
\end{tabular*}
\end{table}
shows the inputs considered for the FT ANN with the corresponding upper and lower bounds.

The overall FT process is characterized by one feed inlet, three distinct product outlet ports (product, water, purge), and two thermal interactions (the exothermic heat release associated with FT synthesis and the heat removal required for post-reaction cooling).

\subsubsection{Linear processes}
For the remaining processes, chemical conversions are modeled based on a simplified linear short-cut model according to \autoref{EQ: 13}. In the following, these processes are described in more detail. Key technical and economic parameters associated with these processes, including their specific energy consumption, operational efficiency, and cost metrics, are provided in Subsection A.3.\@ of the SI. 

\paragraph*{Air separation unit (ASU).}
A cryogenic distillation (CRD)-based ASU is incorporated into the network configuration to provide high-purity \ce{O2} for integration into the gasification stage, thereby improving the quality of the resulting synthesis gas and enhancing overall thermodynamic and process efficiencies. The ASU operates at cryogenic temperatures, typically ranging from approximately \SI{-180}{} to \SI{-196}{\degreeCelsius}, to achieve phase separation of atmospheric air into its principal constituents \cite{Bocker.2010, Palys.2022}. \ce{O2} is recovered as the primary product, while \ce{N2}, which is not utilized within the system boundaries, is treated as a non-value-adding byproduct and subsequently discharged as waste. 

\paragraph*{Direct air capture (DAC).}
Based on Svitnič and Sundmacher \cite{Svitnic.2022}, we consider the Climeworks low temperature DAC process to separate \ce{CO2} from ambient air using electricity for fans and heat for desorption \cite{Climeworks.25.02.2025, Fasihi.2019}.

\paragraph*{Water electrolysis.} \label{Paragraph: Water electrolysis}
The production of high-purity \ce{H2} \emph{via} electrochemical water splitting follows this overall reaction:
\begin{align*}
\ce{H2O} \rightarrow \ce{H2} + 0.5 \, \ce{O2} \quad \Delta_{\mathrm{R}}\tilde{h}^{\theta} = \SI{283.5}{kJ/mol}.
\end{align*}
Not only the \ce{H2} can be used for fuel synthesis, but the produced \ce{O2} can also be utilized as an oxidizing agent in the biomass gasification process \cite{Maggi.2020}. Three distinct electrolysis technologies are included: alkaline electrolysis cells (AEC), proton exchange membrane electrolysis cells (PEMEC), and solid oxide electrolysis cells (SOEC). AEC systems utilize a concentrated aqueous solution of potassium hydroxide (\ce{KOH}) as the electrolyte and incorporate porous diaphragms to separate the anode and cathode compartments. This configuration enables durable and economically favorable operation but is limited by lower achievable current densities and slower dynamic response characteristics. PEMEC systems employ a solid polymer electrolyte, which allows for elevated current densities, rapid transient behavior, and compact system architecture. However, these advantages come at the cost of increased capital expenditure due to the requirement for noble metal catalysts (\emph{e.g.}\@, platinum or iridium). SOEC technology operates under high-temperature conditions (\SI{700}{}–\SI{900}{\degreeCelsius}) and utilizes dense ceramic electrolytes, commonly stabilized zirconia. This thermally assisted operation reduces the electrical energy demand by leveraging external heat sources, thereby enhancing overall efficiency. Nonetheless, the elevated temperatures impose significant material durability and system longevity constraints. Each electrolyzer variant presents specific trade-offs in terms of efficiency, capital intensity, and integration potential, which must be evaluated relative to system objectives and available energy resources \cite{ShivaKumar.2019, Buttler.2018, Palys.2022}.

\paragraph*{Autothermal reforming (ATR).}
To facilitate a techno-economic and environmental comparison between sustainable liquid fuel production pathways and conventional fossil-based routes, this study incorporates an ATR process for the generation of synthesis gas from methane (\ce{CH4}). The ATR process is selected based on its superior thermal efficiency relative to conventional steam methane reforming (SMR), as substantiated by Kim \emph{et al.}\@ \cite{Kim.2021} and Svitnič \emph{et al.}\@ \cite{Svitnic.2024}. Operating under elevated temperatures (approximately \SI{900}{\degreeCelsius}) and moderate to high pressures (\SI{10}{bar} \cite{Kim.2021} to \SI{35}{bar} \cite{Oni.2022}), the ATR process combines partial oxidation and steam reforming within a single reactor, enabling a thermally balanced system. The chemical reactions include:
\begin{align*}
\ce{CH4} + \ce{H2O} &\rightleftharpoons \ce{CO} + \ce{H2} &\;& \Delta_{\mathrm{R}}\tilde{h}^{\theta} = \SI{206.2}{kJ/mol}, \\
\ce{CH4} + 2 \, \ce{O2} &\rightarrow \ce{CO2} + 2 \, \ce{H2O} &\;& \Delta_{\mathrm{R}}\tilde{h}^{\theta} = \SI{-802.3}{kJ/mol}, \\
\ce{CO} + \ce{H2O} &\rightleftharpoons \ce{CO2} + \ce{H2} &\;& \Delta_{\mathrm{R}}\tilde{h}^{\theta} = \SI{-41.2}{kJ/mol}.
\end{align*}
The exothermic partial oxidation of methane provides the thermal energy necessary to drive the endothermic steam reforming reaction, thus enabling autothermal operation. The resultant syngas mixture is characterized by a \ce{H2}-enriched composition relative to \ce{CO}, making it suitable for downstream synthesis processes \cite{Kim.2021}. The \ce{O2} required for oxidation can be supplied \emph{via} CRD or electrolysis, depending on process integration preferences.

\paragraph*{Acid gas removal (AGR).}
Due to the presence of undesirable constituents such as \ce{CO2}, \ce{H2O}, and trace contaminants including \ce{H2S} and \ce{NH3} in the product gas streams generated by biomass gasification and RWGS, a gas purification step is required prior to utilization of these streams as synthesis gas in FT synthesis. To this end, an AGR process based on the Rectisol\textsuperscript{\textregistered} approach is employed, following the design principles outlined by Liu \emph{et al.}\@ \cite{Liu.2015}. In this process, chilled methanol is used as a solvent capable of selectively absorbing acid gases and polar impurities due to its high solubility characteristics at low temperatures. The Rectisol\textsuperscript{\textregistered} technology enables near-complete removal of contaminants such as \ce{H2S}, \ce{NH3}, and \ce{CO2}, thereby producing a high-purity synthesis gas stream suitable for catalytic conversion in FT reactors \cite{Gatti.2014, Liu.2015, Sharma.2016}. The extent of \ce{CO2} removal can be tailored to meet downstream processing requirements.

\paragraph*{Carbon sequestration (CS).}
To incorporate the option of permanent carbon removal within the overall framework, a dedicated CS process is integrated into the superstructure. This process facilitates the transport and subsequent sequestration of high-purity \ce{CO2} streams originating from various upstream processes \cite{Demirhan.2021}. The CS process is modeled to accommodate pure \ce{CO2} as an input and is assumed to achieve complete containment of the captured carbon, thereby contributing to net emissions mitigation within the system boundaries.

\section{Process representation \emph{via} neural networks} \label{Section: Process representation via neural networks}
To address the limitations inherent in conventional superstructure optimization, namely, the reliance on linear input–output representations and the \emph{a priori} fixation of process operating conditions in terms of temperature and pressure, selected subprocesses (gasification, RWGS, FT, as described in \autoref{Subsection: Processes}) within the network are embedded as ANNs. While the incorporation of detailed mechanistic models that account for thermodynamic equilibria, reaction kinetics, and process efficiencies would enhance model fidelity, their direct integration into a MIQCP framework would introduce substantial nonlinearities, thereby impeding the tractability and solvability of the global optimization problem. To circumvent this hurdle, ANNs are trained on data generated from Aspen Plus\textsuperscript{\textregistered} \cite{AspenPlu.24.08.2023} simulations. These surrogate models are subsequently embedded within the optimization framework, enabling the representation of complex nonlinear process behavior while preserving the quadratically constrained nature and associated computational efficiency.

ANNs are extensively employed for data-driven function approximation and have been successfully applied across a broad range of scientific and engineering disciplines \cite{Fernandes.2006, Mujtaba.2006}. Mathematically, a feed-forward ANN is structured as a sequence of $L \in \mathbb{N}$ layers, each performing an affine linear transformation of its input followed by the application of a nonlinear activation function $\sigma: \mathbb{R} \rightarrow \mathbb{R}$, which is typically applied element-wise. The computational operation of each layer $l \in \left[L \right]$ is expressed as:
\begin{align}
x^{(l)} = \sigma^{(l)} \left(W^{(l)} x^{(l-1)} + b^{(l)} \right) \quad \forall l \in \left[L \right],
\end{align}
where $x^{(0)} = x \in \mathbb{R}^{n_x}$ denotes the input vector to the network, and $W^{(l)} \in \mathbb{R}^{n_l \times n_{l-1}}$, $b^{(l)} \in \mathbb{R}^{n_l}$ represent the trainable weight matrix and bias vector associated with layer $l$ \cite{Fischetti.2018}. In this study, the rectified linear unit (ReLU) function is employed as the activation function:
\begin{align}
\mathrm{ReLU}(x) = \text{max}\{0,x\},
\end{align}
which induces a piecewise affine linear mapping within the ANN. This structure renders ReLU-activated networks particularly suitable for integration into MILP formulations. The key principle is that each ReLU neuron can be expressed using linear constraints coupled with a binary decision variable, which encodes the activation status of the neuron -- \emph{i.e.}\@, whether the input is positive (active segment) or non-positive (inactive segment). By explicitly modeling the selection of the active linear segment, a MILP formulation can exactly represent the ANN’s nonlinear mapping.

The integration of ANNs into mathematical optimization frameworks necessitates the introduction of auxiliary decision variables and constraints to replicate the input–output mappings learned by the data-driven surrogate models. A principal advantage of ReLU-activated ANNs lies in their capacity to approximate complex nonlinear relationships while retaining compatibility with MILP formulations through so-called big-$M$ reformulations. These reformulations are widely adopted in MILP to model binary decisions corresponding to activated and non-activated states in piecewise linear functions and can thus be applied to the ReLU activation. Assuming bounded pre-activation inputs for each neuron, \emph{i.e.}\@, $LB_i^{(l)} \leq W_i^{(j)} x^{(l-1)} + b_i^{(l)} \leq UB_i^{(l)}$, the ReLU activation for a single neuron $i$ in layer $l$, given by $x_i^{(l)} = \mathrm{ReLU}\left(W_i^{(l)} x^{(l-1)} + b_i^{(l)}\right)$, can be equivalently represented \emph{via} the following set of MILP constraints:
\begin{align}
x_i^{(l)} &\geq 0, \nonumber \\
x_i^{(l)} &\geq W_i^{(l)} x^{(l-1)} + b_i^{(l)}, \nonumber \\
x_i^{(l)} &\leq W_i^{(l)} x^{(l-1)} + b_i^{(l)} - LB_i^{(l)} \left(1 - \epsilon_i^{(l)} \right), \label{EQ: 63} \\
x_i^{(l)} &\leq UB_i^{(l)} \epsilon_i^{(l)}, \nonumber \\
\epsilon_i^{(l)} &\in \{0,1\}, \nonumber
\end{align}
where $W_i^{(l)}$ denotes the $i$-th row of the weight matrix in layer $l$, and $\epsilon_i^{(l)}$ is a binary variable that activates the appropriate affine segment of the piecewise function \cite{Anderson.2020, Grimstad.2019, Plate.252025}. This formulation enables the exact representation of the ReLU-based ANNs. Consequently, the surrogate models serve as an efficient mechanism for embedding the nonlinear thermodynamic and kinetic process behavior derived from Aspen Plus\textsuperscript{\textregistered} \cite{AspenPlu.24.08.2023} simulations, without exacerbating the overall nonlinearity of the overarching MIQCP formulation described in \autoref{Section: Superstructure optimization with mixtures/model description}.

To facilitate the systematic incorporation of trained ReLU ANNs -- \emph{via} MILP constraints as described in \autoref{EQ: 63} -- into our MIQCP framework, the Optimization \& Machine Learning Toolkit (\texttt{\textbf{OMLT}}), as developed by Ceccon \emph{et al.}\@ \cite{Ceccon.2022}, is employed. \texttt{\textbf{OMLT}} is a Python-based computational toolkit designed for the representation of machine learning models, including neural networks and gradient-boosted decision trees, within deterministic optimization formulations. The package enables the automatic generation of the requisite decision variables and algebraic constraints for embedding trained ANNs within the \texttt{\textbf{Pyomo}} \cite{Bynum.2021, hart2011pyomo} algebraic modeling environment.

\subsection{Artificial neural network training}
The process flowsheets delineated in \autoref{Subsubsection: Processes represented by artificial neural networks} are used to generate the training data sets for the ANNs. To facilitate this, the integrated Aspen Plus\textsuperscript{\textregistered} \cite{AspenPlu.24.08.2023}–Python interface is employed to execute a fully automated simulation workflow. Following the specification of relevant input parameters and their respective bounds, a Latin Hypercube Sampling (LHS) methodology is implemented to ensure a statistically robust and space-filling sampling of the input domain. We explicitly include boundary (extreme) values to enhance model generalizability \cite{Lu.2022, McBride.2019}. A comprehensive enumeration of all output variables associated with the three ANN architectures is provided in Subsection D.1.\@ of the SI.

A distinctive feature of the gasification ANN lies in its capacity to predict the syngas product composition, specifically, the molar fractions of key constituents such as \ce{H2} and \ce{CO}, as a function of the chosen biomass feedstock (see \autoref{Fig.: Aspen Plus flowsheet simulation of gasification process.} and Table C.1.\@ in the SI). To incorporate categorical input data pertaining to biomass type, a one-hot encoding scheme is implemented \cite{Ascher.2022, SanchezMedina.2022}. This encoding methodology transforms discrete categorical variables into binary vectors, wherein each biomass class is uniquely represented by a vector containing a single activated element (value of 1) corresponding to the class identity, with all remaining elements deactivated (value of 0). Such a representation avoids the imposition of artificial ordinal relationships among categories, thereby preserving the nominal nature of the data. Within the gasification ANN framework, this encoding enables the network to differentiate between biomass types and to learn underlying relationships between feedstock characteristics and resulting gas-phase composition.

\begin{figure}[h]
\centering
  \includegraphics[width=0.55\columnwidth]{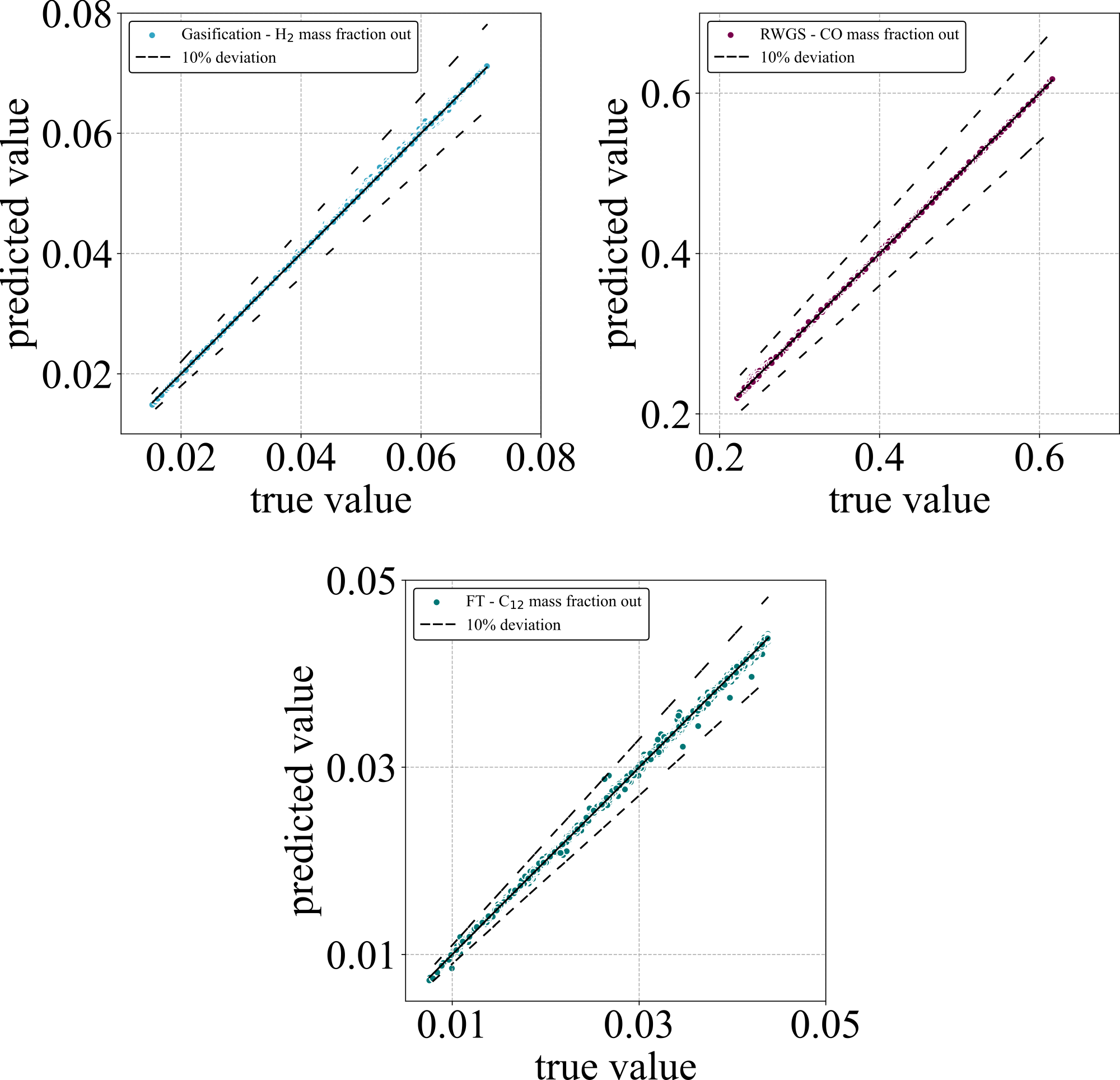}
\caption{Selected parity plots for the artificial neural networks. Test data points are shown for three selected neural network outputs. (\ce{H2} mass fraction exiting the gasification process after the water scrubbing unit; \ce{CO} mass fraction in the output of the reverse water-gas shift process \emph{via} the first outlet port; \ce{C12H26} mass fraction of the Fischer-Tropsch output).}
\label{Fig.: Selected parity performance of the artificial neural networks.}
\end{figure}

The data sets comprise \SI{75000}{} simulation data points per biomass type for the gasification flowsheet, \SI{50000}{} data points for the RWGS process, and \SI{10000}{} data points for the FT synthesis flowsheet. As the primary objective of this study is not to determine the minimal data requirements for accurate ANN training, the data set sizes are selected to be sufficiently large to enable accurate approximation of the underlying process models. All ANNs are implemented and trained using the Keras Application Programming Interface (API) within the TensorFlow framework \cite{Team.10.01.2024}. Prior to training, input and output variables are standardized, and the data is partitioned into training and test subsets following an \SI{80}{}/\SI{20}{} split. Additionally, \SI{20}{\%} of the training subset is reserved for validation during training to monitor generalization and prevent overfitting. Hyperparameter optimization is conducted through systematic tuning procedures, with a full listing of selected hyperparameters provided in Table D.1.\@ in the SI. It is established in the literature that optimization over embedded ANNs becomes significantly more challenging as the number of hidden layers in the ANNs increases \cite{Grimstad.2019, Plate.252025, Tsay.08.02.2021, Schweidtmann.2018}. To achieve an optimal compromise between computational efficiency during integration into the overarching superstructure optimization framework and predictive accuracy, all ANN architectures employ a single hidden layer with varying numbers of ReLU activation neurons. Model training is performed using the Adam optimization algorithm \cite{Kingma.2014}, with mean squared error (MSE) employed as the loss function. A summary of key performance indicators for the trained ANNs is presented in \autoref{Table: Various performance metrics of flowsheet artificial neural networks},
\begin{table}[h]
\small
\caption{Performance metrics of artificial neural networks. $\mathrm{MSE_{train}}$: mean squared error for training data; $\mathrm{MAE}_{\mathrm{test}}$: mean absolute error for test data; $\mathrm{MAPE}_{\mathrm{test}}$: mean absolute percentage error for test data.}
\label{Table: Various performance metrics of flowsheet artificial neural networks}
\begin{tabular*}{1\textwidth}{@{\extracolsep{\fill}}lrrr}
\hline
performance metric & \multicolumn{1}{c}{gasification} & \multicolumn{1}{c}{RWGS} & \multicolumn{1}{c}{FT} \\
\hline
$\mathrm{MSE_{train}}$ & \SI[scientific-notation=true]{1.0760e-4}{} & \SI[scientific-notation=true]{2.8544e-04}{} & \SI[scientific-notation=true]{1.6e-3}{} \\
$R^2_{\mathrm{test}}$ & \SI{0.99994}{} & \SI{0.99992}{} & \SI{0.99955}{} \\
$\mathrm{MAE}_{\mathrm{test}}$ & \SI{0.000453}{} & \SI{0.000497}{} & \SI{0.000315}{} \\
$\mathrm{MAPE}_{\mathrm{test}}$ & \SI{0.004120}{} & \SI{0.001713}{} & \SI{0.064073}{} \\
\hline
\end{tabular*}
\end{table}
while representative parity plots illustrating prediction accuracy for each ANN are shown in \autoref{Fig.: Selected parity performance of the artificial neural networks.}.
It is evident that the ANN architectures with only one hidden layer per network are sufficient to predict the relationships between the inputs and outputs with satisfactory accuracy. Further details regarding network training and performance evaluation are given in Subsection D.3.\@ of the SI.

\subsection{Embedding} \label{Subsection: Superstruture optimization with embedded artificial neural networks}
Following the training of the ANNs, the corresponding decision variables and algebraic constraints necessary for representing the network architecture and input–output behavior are integrated into the overall MIQCP formulation \emph{via} the \texttt{OMLT} framework \cite{Ceccon.2022}, as detailed in \autoref{Section: Superstructure optimization with mixtures/model description}. This integration involves establishing equality constraints that directly couple the ANN input and output variables (denoted as $f^{\mathrm{ANN}}_{j,\mathrm{(input,}i)}$ and $f^{\mathrm{ANN}}_{j,\mathrm{(output,}o)}$ for process $j$ and input/output $i$/$o$) to the decision variables defined within the superstructure (see \autoref{Fig.: General process modeling.}). For instance, the decision variable representing the reaction temperature in the biomass gasification step is mapped to the seventh input node of the corresponding gasification ANN ($T_{\mathrm{(Gasification,2)}} = f^{\mathrm{ANN}}_{\mathrm{Gasification,(input,7)}}$). Output variables of the ANNs are then expressed as functions of these linked inputs. As an illustrative example, the mass fractions of hydrocarbon species at the first outlet port of the FT synthesis process are computed \emph{via} the trained FT ANN model according to the following relation:
\begin{align}
w^{\mathrm{out}}_{\mathrm{C_{n}H_{2n+2},(FT,1)}} &= y_{\mathrm{FT}} f^{\mathrm{ANN}}_{\mathrm{FT,(output,n+2)}} \quad \forall \mathrm{n} = 1, \,  ..., \, 30.
\end{align}
Important operating parameters such as the reaction temperature, the pressure and the mass fraction of \ce{H2} at the inlet ($T_{\mathrm{(FT,1)}}, \, p_{\mathrm{FT}}, \, w^{\mathrm{in}}_{\mathrm{H_2,(FT,1)}}$) are taken into account as ANN inputs and thus influence the outputs. Through this formulation, all input and output variables of the embedded ANNs are hard-linked to the superstructure’s decision variables, ensuring full integration of data-driven surrogate models within the global optimization framework. Subsection D.4.\@ in the SI provides a complete list of the associated equality and inequality constraints.

In order to represent the categorical inputs of the gasification ANN, the binary variables used to describe the biomass types (introduced in \autoref{Paragraph: Biomass gasification}) are mapped to the first three input nodes of the ANN, such that each input node is constrained to assume binary values (\SI{0}{} or \SI{1}{}). In this way, the one-hot encoding paradigm is adhered to and it is ensured that at most one biomass type is active.  

The integration of ANNs within the optimization framework facilitates the direct inclusion of both ANN inputs and outputs as decision variables in the higher-level MIQCP formulation. This bidirectional embedding allows for the imposition of explicit constraints on both the input and output domains of the ANNs, thereby enabling optimization not only in the forward predictive direction but also in reverse. Consequently, the introduction of supplementary decision variables at the plant- and unit-operation level substantially broadens the design space and increases the degrees of freedom (\emph{e.g.}\@, reaction temperature, inlet composition).

All instances of the MIQCP problem incorporating embedded ANNs are solved on an Apple MacBook Pro equipped with an M4 Max system-on-chip, comprising a 16-core CPU, a 40-core GPU, and a unified memory architecture with a total of 48 GB of RAM. The optimization tasks are executed \emph{via} the \texttt{\textbf{Gurobi}} Optimizer \cite{GurobiOptimization.05.10.2023}, version 12.0.1, interfaced through the \texttt{\textbf{Pyomo}} \cite{Bynum.2021, hart2011pyomo} modeling framework. The full optimization formulation encompasses approximately \SI{6000}{} continuous and \SI{250} binary decision variables, of which \SI{200} binary variables specifically encode the activation patterns of ReLU-based hidden neurons within the embedded ANNs. By leveraging warm-start initialization techniques -- utilizing solutions obtained from prior optimization runs --, and methodological enhancements outlined in Plate \emph{et al.}\@ \cite{Plate.252025}, computational solution times are consistently maintained below \SI{500}{s}.

\section{Results} \label{Section: Results}
\subsection{Cost vs. \ce{CO2} emissions} \label{Subsection: Cost vs. CO2 emissions}
As detailed in \autoref{Subsubsection: Economic constraints}, the principal objective of the optimization framework is the derivation of cost-optimal process configurations for the production of the kerosene fraction. To incorporate both economic performance and environmental sustainability, the optimization is extended to yield Pareto-optimal solutions that simultaneously account for \ce{CO2} emissions arising from both the production and end-use combustion. The multi-objective optimization problem is addressed \emph{via} the $\epsilon$-constraint method \cite{Yv1971OnAB}, wherein the \ce{CO2} emissions, quantified according to the mass-balances in \autoref{Subsubsection: CO2 balancing}, are systematically constrained. The minimization of annualized CAPEX and OPEX is retained as the primary objective function.

\begin{figure*}
\centering
\includegraphics[width=0.9\columnwidth]{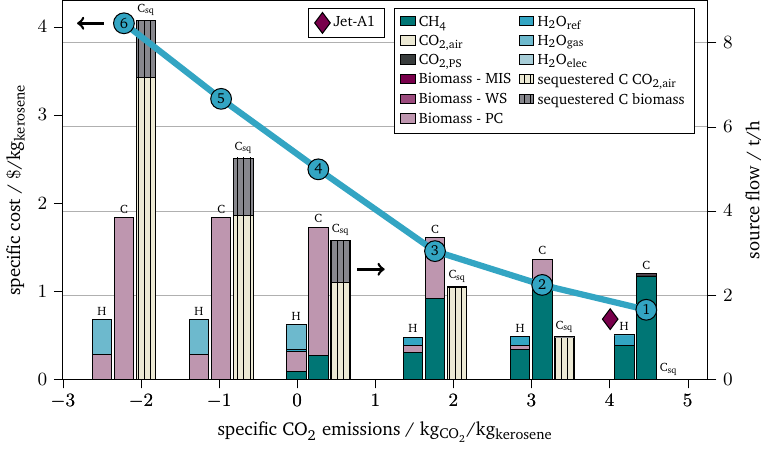}
\caption{Pareto-optimal designs for specific kerosene cost vs. specific \ce{CO2} emissions (left y-axis). The bars represent the source flows for carbon (C) and hydrogen (H) entering the system (right y-axis). The third bar shows sequestered carbon (C\textsubscript{sq}) either from atmospheric \ce{CO2} or biomass. As a reference, typical Jet-A1 aviation fuel is shown as a red diamond. (PS: point source; MIS: miscanthus; WS: wheat straw; PC: pine chips; ref: reforming; gas: gasification; elec: electrolysis).}
\label{Fig.: Cost vs. CO2 results}
\end{figure*}

\autoref{Fig.: Cost vs. CO2 results} presents the resulting Pareto frontier, depicted in blue, \emph{i.e.}\@, the specific kerosene production costs (left y-axis) as a function of the corresponding specific \ce{CO2} emissions.
Here, “specific” denotes normalization with respect to the mass flow rate of kerosene exiting the FT synthesis stage under the respective optimized operating conditions. Additionally, the right y-axis displays the elemental feedstock flows for both C and H, corresponding to each Pareto-optimal configuration. A third bar is included to indicate the amount of sequestered carbon (denoted as C\textsubscript{sq}), originating either from DAC or from biogenic sources. Cost and carbon footprint of conventional Jet-A1 aviation fuel are included for comparison \cite{Salem.2023, Vo.2024}.

\subsubsection{Least-cost reference design (autothermal reforming pathway)}
The least-cost design configuration emerges under conditions without constraints on \ce{CO2} emissions, when synthesis gas is generated exclusively \emph{via} the ATR process (Pareto-optimal design number 1 in \autoref{Fig.: Cost vs. CO2 results}). In this scenario, fossil-derived natural gas serves as the primary C and H source for the downstream production of the kerosene fraction. Due to the inherently hydrogen-rich composition of syngas produced \emph{via} ATR, a supplementary RWGS unit of limited capacity is integrated to increase the \ce{CO} content by converting externally sourced \ce{CO2} (originating from concentrated point sources) with the surplus \ce{H2}. This \ce{H2} originates both from the natural gas feedstock and the steam introduced during the reforming reactions. The resulting ATR-based reference configuration yields a specific production cost of approximately \SI{0.79}{\$/kg\textsubscript{kerosene}}, and specific \ce{CO2} emissions of \SI{4.47}{kg\textsubscript{\ce{CO2}}/kg\textsubscript{kerosene}}, accounting for both process-related (\SI{1.37}{kg\textsubscript{\ce{CO2}}/kg\textsubscript{kerosene}}) and end-use combustion emissions (\SI{3.10}{kg\textsubscript{\ce{CO2}}/kg\textsubscript{kerosene}}). Notably, the analysis does not consider further upgrading of FT products (\emph{e.g.}\@, hydrocracking), and heavier hydrocarbon fractions cannot be converted to kerosene within the current system boundaries. Total emissions are allocated to all fractions based on energy content (see \autoref{Subsubsection: Post optimization calculations}). Upgrading could increase the yield at the cost of additional processes; we therefore may be slightly overestimating the specific cost and emissions. Nevertheless, the cost and emission metrics for the ATR reference case agree with reported numbers for fossil-derived kerosene in existing literature \cite{Salem.2023, Vo.2024}.

\subsubsection{Net zero and carbon-negative configurations}
Progressively constraining \ce{CO2} emissions results in a systematic shift in process design and a considerable increase in the specific cost of kerosene. Near the net-zero emissions point, the most cost-effective pathway is a hybrid configuration that combines ATR with biomass gasification (Pareto-optimal design number 4 in \autoref{Fig.: Cost vs. CO2 results}), where the majority of the C and \ce{H} originate from biomass. In this case, kerosene production costs about \SI{2.38}{\$/kg\textsubscript{kerosene}}, which is roughly three times higher than in the unrestricted reference system. In the extreme case, characterized by negative specific \ce{CO2} emissions (Pareto-optimal design number 6 in \autoref{Fig.: Cost vs. CO2 results}), the kerosene production cost reaches \SI{4.04}{\$/kg\textsubscript{kerosene}}, representing an increase by a factor greater than five relative to the ATR-based reference system. Under this configuration, synthesis gas for the FT process is generated exclusively \emph{via} gasification of biomass, specifically pine wood chips, thereby eliminating reliance on fossil feedstocks and substantially reducing net GHG emissions. By combining kerosene production with the removal of atmospheric and biogenic carbon, it is essentially possible to produce carbon-negative SAF.

\autoref{Fig.: Cost vs. CO2 results} elucidates that the observed reduction in specific \ce{CO2} emissions associated with kerosene production is governed by two principal mechanisms. First, there is a progressive substitution of fossil-derived \ce{CH4} with biomass as the primary C and H source. Along the continuum of the Pareto frontier, spanning from the cost-optimal (design number 1) to the most carbon-negative design (design number 6), hybrid process configurations emerge in which synthesis gas is co-generated \emph{via} both ATR and biomass gasification in varying proportions. This compositional shift in syngas production exerts a dual impact on process-related GHG emissions. Substitution of \ce{CH4} reduces upstream supply chain emissions associated with the fossil feedstock. Additionally, the biogenic origin of the carbon in the biomass reduces the carbon footprint, as atmospheric \ce{CO2} is taken up during feedstock growth. Nevertheless, the intrinsic carbon neutrality of biomass alone is insufficient to achieve net zero or net-negative \ce{CO2} emissions across the system boundary. This is in part due to the carbon footprint of electricity and heat as well as the supply chain emissions associated with the biomass. To attain further emission abatement, carbon sequestration is incrementally deployed. This system actively stores \ce{CO2}, thereby preventing its release into the atmosphere. The vertically striped bars in \autoref{Fig.: Cost vs. CO2 results} denote the sources of sequestered carbon, revealing a notable optimization outcome: the model predominantly favors the integration of DAC technology for atmospheric \ce{CO2} extraction and subsequent storage. Any leftover \ce{CO2} from biomass gasification is sequestered. The majority to fulfill the sequestration requirement is then covered by DAC-CS. This outcome reflects an economically rational allocation of carbon flows, prioritizing the utilization of biomass-derived carbon for synthetic fuel production, while relying on DAC to meet constraints on net \ce{CO2} emissions.

In the absence of dedicated carbon sequestration, the mitigation of combustion emissions can be achieved solely through the biogenic carbon uptake intrinsic to biomass utilization or \emph{via} the incorporation of atmospheric \ce{CO2} through DAC. However, under the parameter assumptions employed in this study, the minimum attainable specific \ce{CO2} emissions, without CS, are limited to \SI{3.08}{kg\textsubscript{\ce{CO2}}/kg\textsubscript{kerosene}}. While this value represents a reduction relative to the emissions profile of conventional Jet-A1 fuel, it remains significantly above the net zero emissions threshold. Any subsequent reduction in the overall \ce{CO2} footprint requires the presence of CS. The principal constraint preventing further emission reductions in the non-sequestration scenario arises from GHG emissions associated with the external supply of biomass, electricity and thermal energy. If a hypothetical scenario is considered in which the provision of electricity and heat is entirely decarbonized (\emph{i.e.}\@, associated with zero \ce{CO2} emissions) the net zero target of \SI{0}{kg\textsubscript{\ce{CO2}}/kg\textsubscript{kerosene}} becomes theoretically attainable. In this idealized case, atmospheric \ce{C} is sourced \emph{via} DAC, and \ce{H} demand is met through water electrolysis powered by renewable electricity. Notably, reliance on biomass gasification alone is insufficient to achieve net zero emissions due to residual supply chain emissions associated with the harvesting, processing, and transport of various biomass feedstocks. However, if burden-free biomass (\emph{i.e.}\@, without supply chain emissions) were available, net zero could also be reached at considerably lower costs than in the DAC-based pathway, which -- despite being technically feasible -- remains associated with substantially higher production costs. Consequently, a system with net zero emissions necessitates either a substantial reduction of supply chain emissions -- biomass, electricity, heat -- or the systematic integration of carbon sequestration technologies into the fuel production.

A more granular examination of the transition between the intermediate configurations on the Pareto frontier, specifically between the third and fourth points in \autoref{Fig.: Cost vs. CO2 results}, highlights the benefits of adaptively optimizing the operation of individual subprocesses in response to increasing \ce{CO2} emission constraints. This analysis reveals a progressive substitution of fossil-derived natural gas with lignocellulosic biomass, particularly pine chips, as the principal feedstock. The preferential selection of pine chips over alternative biomass types is driven by their comparatively low OPEX and their high carbon content, as detailed in Table C.1.\@ in the SI. To enable enhanced biogenic carbon sequestration and compliance with more rigorous emission thresholds, the gasification process is reconfigured to produce outlet gas streams with significantly elevated \ce{CO2} mass fractions. From the fourth Pareto point onward, the syngas exiting the gasifier contains at least \SI{47}{wt.\%} \ce{CO2}, in contrast to the \textapprox\SI{6}{wt.\%} \ce{CO2} observed at earlier Pareto points (design numbers 1-3). The \ce{CO2} generated in this regime is no longer utilized for synthetic fuel synthesis, it is instead sequestered to reduce net emissions. The increase in \ce{CO2} yield from gasification is facilitated by enhanced oxidation of the carbonaceous feedstock, necessitating a higher oxygen-to-biomass feed ratio. A CRD subsystem is deployed to supply the requisite additional \ce{O2}. As a result, the gasification process transitions from an oxygen-lean to an oxygen-rich regime. Specifically, the oxygen-to-biomass ratio increases from approximately \SI{0.05}{} at the third Pareto point (representing the lower operational boundary) to nearly \SI{0.3}{} at the fourth point, underscoring the substantial process adaptation required to meet low-emission targets through integrated carbon management.

The results depicted in \autoref{Fig.: Cost vs. CO2 results} further underscore that the utilization of renewable carbon sources (specifically biomass) is essential for achieving net zero \ce{CO2} emissions in the production of SAFs. In the absence of biomass-based feedstocks, the optimization process favors configurations wherein synthesis gas is generated exclusively \emph{via} the fossil-based ATR pathway, and \ce{CO2} emissions are mitigated by DAC-CS. While such designs are technically capable of meeting emission constraints, they exhibit reduced economic efficiency relative to hybrid configurations incorporating biomass gasification. For instance, at the fourth Pareto-optimal point, approaching net zero emissions, a configuration relying solely on ATR-based syngas production incurs a specific kerosene production cost of \SI{2.65}{\$/kg\textsubscript{kerosene}}. In contrast, a mixed feedstock design that integrates both biomass and fossil natural gas yields a lower specific cost of \SI{2.38}{\$/kg\textsubscript{kerosene}}, representing an approximate \SI{11}{\%} cost reduction. These findings indicate that, from both an economic and environmental perspective, biomass-enabled process routes offer a superior compromise for SAF production. Note that, as alluded to earlier, biomass supply chain emissions are assumed to be relatively high. When biomass with lower carbon footprint becomes available, biomass-based designs become more favorable compared to the fossil reference.

\subsubsection{Comparative analysis of carbon supply pathways}
\begin{figure}[h]
\centering
\includegraphics[width=0.75\columnwidth]{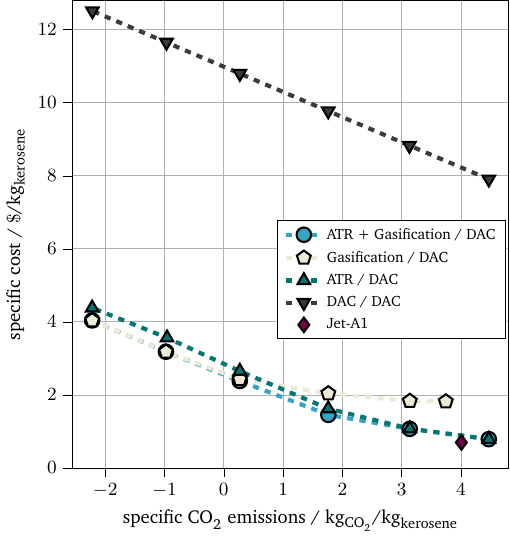}
\caption{Pareto-optimal designs for specific kerosene cost vs. specific \ce{CO2} emissions for different carbon sources - fossil (ATR), biogenic (gasification), atmospheric (DAC). “/ DAC” indicates that all designs require carbon sequestration to meet emissions targets. Jet-A1 aviation fuel is included for comparison. (ATR: autothermal reforming; DAC: direct air capture).}
\label{Fig.: Comparison of Pareto-optimal designs for specific kerosene cost vs. specific CO2 emissions}
\end{figure}

The conclusions from \autoref{Fig.: Cost vs. CO2 results} are further substantiated by the comparative analyses presented in \autoref{Fig.: Comparison of Pareto-optimal designs for specific kerosene cost vs. specific CO2 emissions}, which
illustrates multiple Pareto-optimal frontiers corresponding to different carbon supply configurations: ATR+Gasification (blue), Gasification-only (beige), ATR-only (green), and DAC-only (black). In all cases, low \ce{CO2} emission targets are achievable only \emph{via} DAC-CS, which is indicated by the “/ DAC”. Among these cases, the ATR+Gasification scenario (blue curve) consistently exhibits the lowest specific production costs across the full range of specific \ce{CO2} emissions. This curve corresponds to the fully unconstrained reference Pareto front previously shown in \autoref{Fig.: Cost vs. CO2 results}. Excluding the ATR process yields the Gasification-only case (beige curve), where biomass serves as the sole carbon source. While this configuration is less favorable than the blue and green cases at higher emission levels due to the relatively high capital and operational costs of biomass conversion, it becomes economically superior under emission constraints. At near net zero emissions, this biomass-based configuration achieves specific kerosene production costs of \SI{2.43}{\$/kg\textsubscript{kerosene}}, approximately \SI{9}{\%} lower than the corresponding ATR-only design reliant on DAC-CS (green curve, \SI{2.65}{\$/kg\textsubscript{kerosene}}). This confirms the economic and environmental advantages of integrating biomass gasification in SAF pathways targeting low emission levels. In contrast, the ATR-only case (green curve), which excludes biomass gasification, depends entirely on fossil-derived synthesis gas, with net emissions offset \emph{via} DAC-CS. While production costs are comparable to the unconstrained case at high emission levels, the cost penalty increases significantly as emission constraints tighten, underscoring the diminishing economic viability of fossil-based pathways when combined with DAC-CS as the sole mitigation strategy. Finally, the DAC-only scenario (black curve) excludes both ATR and gasification, thereby relying exclusively on atmospheric \ce{CO2} \emph{via} DAC as the carbon source, and water electrolysis for \ce{H2}. This pathway exhibits the highest production costs across all emission levels, reaching up to \SI{10.8}{\$/kg\textsubscript{kerosene}} near net zero conditions. This substantial economic burden renders the DAC-only route uncompetitive relative to configurations incorporating natural gas or biomass. The findings underscore that cost-effective and climate-aligned SAF production depends on introducing C into the system in a thermodynamically favorable form, such as biomass or \ce{CH4}, rather than as fully oxidized \ce{CO2}. In addition, the carbon footprint of electricity plays a major role, as these supply chain emissions need to be offset by additional DAC. In summary, bio-based and fossil-plus-offset configurations yield comparable production costs at low emissions, with the biomass-based pathway being slightly more economical near net zero (\SI{2.43}{\$/kg\textsubscript{kerosene}} vs. \SI{2.65}{\$/kg\textsubscript{kerosene}} for ATR+DAC). In contrast, pathways relying solely on atmospheric carbon (DAC-only) result in substantially higher costs and remain far from being competitive, highlighting the practical limitations of producing SAF from \ce{CO2} that enters the system in an energetically less favorable form.

\subsubsection{\ce{CO2} abatement costs and policy implications}
The \ce{CO2} abatement costs presented in this study are calculated relative to the fossil-based ATR configuration (no restrictions on \ce{CO2} emissions), which serves as the reference design (Pareto-optimal design number 1 in \autoref{Fig.: Cost vs. CO2 results}). Applying the \ce{CO2} abatement cost formulation in \autoref{EQ: 59} to the gasification-only configuration at near net-zero emissions yields an estimated abatement cost of approximately \SI{390}{\$/t\textsubscript{\ce{CO2}}}. This value is comparable to previously reported ranges in the literature for SAF pathways with similar technology readiness levels \cite{GonzalezGaray.2022, Salem.2023}. From a policy standpoint, this benchmark indicates that a carbon price of roughly \SI{390}{\$/t\textsubscript{\ce{CO2}}} would be required to incentivize deployment of biomass-based SAF production over fossil-derived alternatives.

Extending this analysis to the two extreme points on the Pareto front shown in \autoref{Fig.: Cost vs. CO2 results}, namely the fossil-based ATR baseline and the biomass-only configuration with net-negative emissions of \SI{-2.22}{kg\textsubscript{\ce{CO2}}/kg\textsubscript{kerosene}}, yields an abatement cost of \SI{485}{\$/t\textsubscript{\ce{CO2}}}, which confirms that abatement costs can also be consistently defined for negative-emission SAF designs. A detailed breakdown of \ce{CO2} abatement costs across intermediate points along the Pareto curve is provided in Table E.1.\@ in the SI. By contrast, process routes relying exclusively on electrolysis-based \ce{H2} and DAC-derived \ce{CO2} exhibit much higher specific production costs of \SI{10.8}{\$/kg\textsubscript{kerosene}} at near-zero emissions, corresponding to abatement costs of approximately \SI{2380}{\$/t\textsubscript{\ce{CO2}}}. These results highlight not only the significant economic advantage of biomass-based configurations but also their markedly higher cost-effectiveness in terms of avoided emissions compared to DAC-dependent pathways. Nevertheless, the viability of biomass-based approaches must be evaluated within the broader context of intersectoral competition for biomass resources, including demand from the energy, agricultural, and industrial sectors \cite{Beer.2025, Svitnic.2024}. To capture the implications of such supply constraints, additional Pareto-optimal process configurations were derived under varying biomass availability scenarios. These scenarios elucidate the sensitivity of outcomes to biomass feedstock limitations and highlight the diminishing cost advantage of biomass-based production as resource availability declines. Comprehensive Pareto-optimal design results under biomass supply constraints are provided in Subsection E.2.\@ of the SI.

\subsubsection{Sensitivity to key parameters}
The results shown in this section are highly dependent on the parameters used, particularly with regard to \ce{CO2} footprints of electricity, heat, and biomass. Table E.2.\@ in the SI contains further information on emissions and costs of plant designs with varying net zero target, supply chain emissions, biomass carbon footprint, and availability of carbon sequestration.

\subsection{Cost breakdown}
\begin{figure}[h]
\centering
\includegraphics[width=0.64\columnwidth]{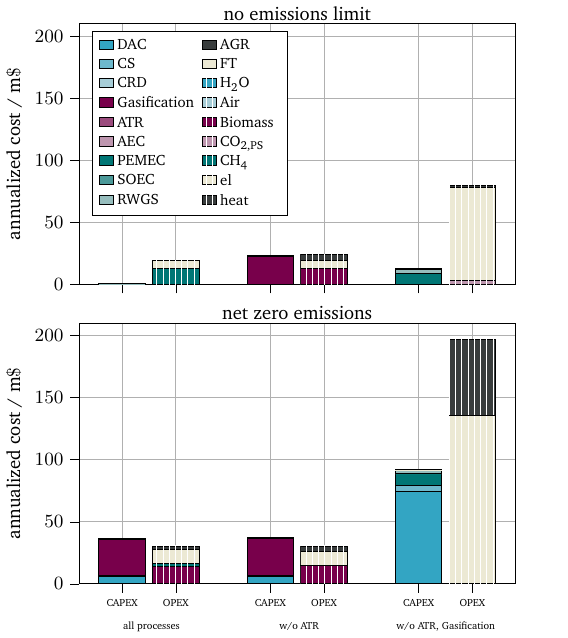}
\caption{Breakdown of annual CAPEX and OPEX for kerosene production designs. The top diagram shows the cost breakdowns for three different cases without restrictions on \ce{CO2} emissions. On the far left design, no processes are excluded in advance, while in the other two cases first the autothermal reforming (ATR) process and finally both the ATR and the gasification process are excluded. The diagram below shows the cost breakdowns for the same three cases, but the greenhouse gas (GHG) emissions are restricted to approximately \SI{0}{kg\textsubscript{\ce{CO2}}/kg\textsubscript{kerosene}}. (DAC: direct air capture; CS: carbon sequestration; CRD: cryogenic distillation; ATR: autothermal reforming; AEC: alkaline electrolyzer; PEMEC: proton-exchange membrane electrolyzer; SOEC: solid-oxide electrolyzer; RWGS: reverse water-gas shift; AGR: acid gas removal ; FT: Fischer-Tropsch; PS: point source; el: electricity).}
\label{Fig.: CAPEX OPEX results}
\end{figure}

\autoref{Fig.: CAPEX OPEX results} presents annualized CAPEX and OPEX across three design scenarios.
These include: (i) an unconstrained configuration wherein all synthesis pathways are available; (ii) a scenario excluding the ATR process ("w/o ATR"); and (iii) a configuration in which both the ATR and biomass gasification routes are omitted ("w/o ATR, Gasification"). The top part of the figure depicts the costs under unconstrained \ce{CO2} emissions, reflecting purely cost-optimal process allocations. In contrast, the bottom part imposes a constraint of \SI{0}{kg\textsubscript{\ce{CO2}}/kg\textsubscript{kerosene}}.

An examination of the top diagram in \autoref{Fig.: CAPEX OPEX results}, corresponding to the scenario without \ce{CO2} emission constraints, reveals that, under the configuration with all processes inlcuded, the process design is exclusively reliant on ATR for synthesis gas production. In this configuration, OPEX, primarily driven by \ce{CH4} as feedstock, constitute the dominant share of the total annualized costs. Upon exclusion of the ATR process, a substantial increase in annual CAPEX is observed, attributable to the capital-intensive biomass gasification infrastructure. In the scenario wherein both the ATR and gasification routes are omitted, the \ce{H2} required for FT synthesis must be supplied entirely \emph{via} water electrolysis. Concurrently, \ce{CO} is generated by catalytically activating captured \ce{CO2} through the RWGS reaction, necessitating additional \ce{H2} input. In the absence of carbon constraints, the use of \ce{CO2} from concentrated point sources emerges as the most cost-effective carbon feedstock. As indicated in \autoref{Fig.: CAPEX OPEX results}, the pronounced increase in total system costs for electrolysis-based process routes is caused by the substantial electrical energy demand.

The bottom diagram in \autoref{Fig.: CAPEX OPEX results} reflects scenarios constrained to approximately net zero \ce{CO2} emissions. In the case of the far left configuration, biomass is employed as the principal \ce{C} and \ce{H} source for synthesis gas generation, with fossil \ce{CH4} contributing only marginally. In this case, annualized CAPEX and OPEX are nearly equivalent. Exclusion of the ATR process exerts minimal influence on the overall process configuration, as its installed capacity in the first case is already marginal. A comparative assessment of the “w/o ATR” scenarios between the top (no emissions limit) and bottom (net zero emissions) diagrams reveals that the primary distinction lies in the implementation of a DAC–CS combination in the bottom case, which is necessary to meet GHG reduction targets through atmospheric carbon removal. In contrast, achieving net zero emissions without the utilization of biomass (\emph{i.e.}\@, under the dual exclusion of both ATR and gasification pathways) incurs a dramatic increase in both CAPEX and OPEX. In such a configuration, all \ce{H2} required for FT synthesis must be generated \emph{via} water electrolysis, resulting in exceptionally high electricity demand, similar to that observed in the unconstrained scenario. However, unlike the top diagram where point-source \ce{CO2} remains viable, the net zero constraint necessitates atmospheric \ce{CO2} exclusively \emph{via} DAC. This not only increases the electrical energy requirement but also introduces significant thermal energy demand for the desorption stage of the DAC system. Consequently, the combined effect of intensive electricity consumption for electrolysis and the thermal requirements of DAC culminates in exceptionally elevated total annualized costs, rendering the biomass-free, net zero SAF production route economically disadvantageous under the given assumptions.

\subsection{Fixed vs. adaptable design} \label{Subsection: Fixed vs. adaptable design}
In most superstructure optimization studies, process variables (\emph{e.g.}\@, reactor temperature, pressure, inlet composition) are fixed prior to optimization \cite{Demirhan.2021, Ganzer.2020, GonzalezGaray.2022, Niziolek.2017, Svitnic.2022}, which implies that individual subprocesses cannot respond to varying external conditions such as changing energy prices or resource availability. As a result, important adjustment potentials remain unexploited. The benefits of allowing the subprocesses to adjust have been previously demonstrated in \autoref{Subsection: Cost vs. CO2 emissions}. These advantages are made accessible only through the integration of ANNs into the superstructure optimization framework, which enables the inclusion of varying operating modes and decision variables at both the plant wide and individual process unit levels. To further elucidate the impact of incorporating these additional degrees of freedom, a sensitivity analysis is performed comparing two scenarios: an "adaptable case", in which the input and output parameters of the embedded ANNs are optimized to respond to changing external conditions, and a "fixed case", where process parameters such as pressure, temperature, and inlet compositions are predetermined and held constant throughout the optimization. For this analysis, the ATR process is excluded \emph{a priori}, and no constraints are imposed on \ce{CO2} emissions.

\begin{figure}[h]
\centering
\includegraphics[width=0.55\columnwidth]{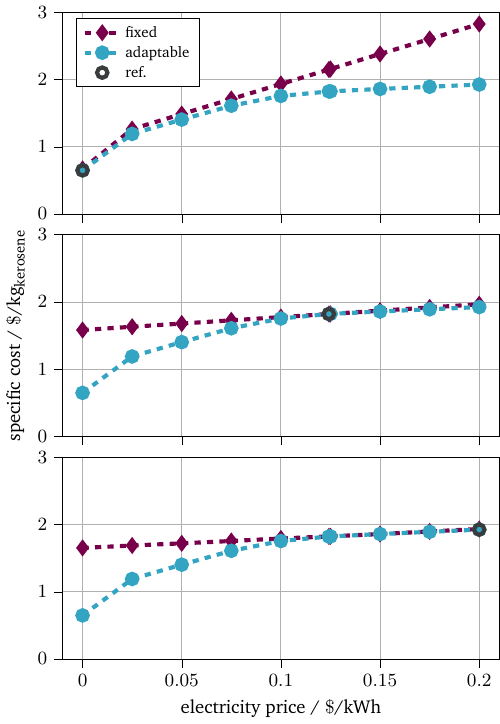}
\caption{Results of the sensitivity analysis for the specific kerosene production costs as a function of the electricity price for the adaptable (blue) and the fixed case (red). In the adaptable case, all decision variables at plant and process unit level (\emph{e.g.}\@, pressure, temperature, inlet compositions) of the processes that are represented \emph{via} ANNs are taken into account. In the fixed case, these degrees of freedom are fixed in advance, depending on the reference case (ref.), and are not part of the actual optimization. All results are generated excluding the autothermal reforming (ATR) process and there are no restrictions regarding \ce{CO2} emissions.}
\label{Fig.: Results fixed vs. adaptable design}
\end{figure}

\autoref{Fig.: Results fixed vs. adaptable design} presents the resulting specific kerosene production costs as a function of electricity price, varied within the range of \SI{0}{\$/kWh} to \SI{0.2}{\$/kWh}.
The blue curves correspond to the adaptable case, incorporating multiple operational decision variables at the plant and process unit levels (\emph{e.g.}\@, variations in pressure, temperature, and inlet compositions). In contrast, the red curves represent the fixed case, where these additional degrees of freedom are omitted. For comparative purposes, selected operating points from the adaptable case are used to define the fixed process conditions, indicated as black unfilled reference markers in \autoref{Fig.: Results fixed vs. adaptable design}. Specifically, the top, middle, and bottom subplots show the fixed case using process conditions derived from the adaptable solutions at electricity prices of \SI{0}{\$/kWh}, \SI{0.124}{\$/kWh}, and \SI{0.2}{\$/kWh}, respectively.

The results reveal a distinct nonlinear relationship between the specific cost of kerosene production and electricity price. In the adaptable case, costs range from approximately \SI{0.65}{\$/kg\textsubscript{kerosene}} at \SI{0}{\$/kWh} to \SI{1.92}{\$/kg\textsubscript{kerosene}} at \SI{0.2}{\$/kWh}, with diminishing sensitivity at higher electricity prices. Across all subplots, the adaptable case consistently shows lower costs than the fixed case. In several instances, the adaptable configuration achieves more than \SI{20}{\%} cost savings relative to the fixed scenario. These findings highlight the critical value of simultaneously optimizing system-level design and process-level operating conditions.

\begin{figure}[h]
\centering
\includegraphics[width=0.55\columnwidth]{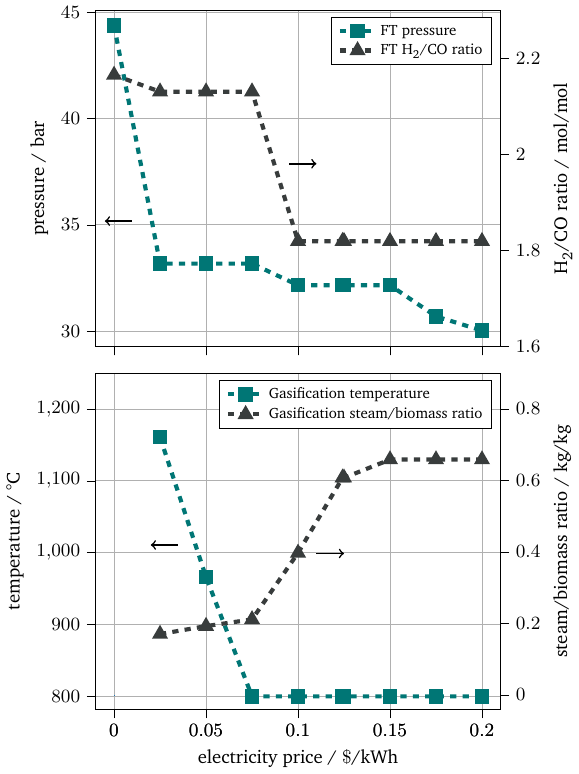}
\caption{Selected optimal operating parameters of the gasification and Fischer-Tropsch (FT) processes as a function of the electricity price. The top diagram shows the pressure of the FT process (green) and the molar \ce{H2}/CO ratio in the inlet (black). The bottom diagram illustrates the temperature of the biomass gasification (green) and the mass-based steam to biomass ratio (black).}
\label{Fig.: Operating parameters results}
\end{figure}

To elucidate the mechanisms underlying the superior economic performance of the adaptable case, a detailed evaluation of the optimal operating conditions for the biomass gasification and FT synthesis processes as a function of electricity price is presented. The top diagram of \autoref{Fig.: Operating parameters results} illustrates the optimal FT reactor pressure (left y-axis) and the corresponding molar \ce{H2}/\ce{CO} ratio at the reactor inlet (right y-axis).
The bottom diagram depicts the gasification temperature (left y-axis) and the mass-based steam-to-biomass ratio (right y-axis). In the hypothetical scenario where electricity is available at no cost (\SI{0}{\$/kWh}), the synthesis gas is entirely derived from water electrolysis in conjunction with \ce{CO2} activation \emph{via} the RWGS reaction. Under such conditions, biomass gasification is rendered superfluous and thus omitted from the process configuration. The favorable \ce{H2} availability allows operation of the FT synthesis at an elevated inlet \ce{H2}/\ce{CO} molar ratio of approximately \SI{2.16}{}. However, a high \ce{H2}/\ce{CO} ratio shifts the ASF product distribution towards lower molecular weight hydrocarbons. To counteract this effect and steer the hydrocarbon product distribution back towards the desired kerosene-range fraction (\ce{C8}–\ce{C16}), the FT reactor is operated at an increased pressure (approximately \SI{44}{bar}), a condition made feasible by the negligible cost of electricity. As electricity prices increase, a transition in both process configuration and operating strategy is observed. The synthesis gas supply gradually shifts from electrolysis and RWGS to biomass gasification. At moderately low electricity prices, a hybrid strategy is employed wherein gasification supplements electrolysis. In this regime, gasification is optimized for high carbon monoxide yield, favoring elevated temperatures and reduced steam-to-biomass ratios. However, once the electricity price exceeds approximately \SI{0.075}{\$/kWh}, the economic viability of electrolysis-derived \ce{H2} diminishes significantly. At this point, biomass gasification must solely meet the requirements of the FT synthesis. To achieve a synthesis gas composition conducive to high kerosene yields, the gasifier operating temperature is reduced and the steam-to-biomass ratio is increased, promoting enhanced \ce{H2} formation—up to \SI{7}{\%} by mass in the product gas. With rising electricity prices, two additional parallel trends emerge. First, syngas compression becomes a major contributor to OPEX, making it optimal to reduce the FT reactor pressure to approximately \SI{30}{bar}. At such lower pressures, product selectivity shifts toward lighter hydrocarbons. Second, the optimal \ce{H2}/\ce{CO} molar ratio decreases to around \SI{1.82}{}, reflecting the reduced hydrogen content in the syngas as a consequence of high electricity-driven hydrogen costs. Because these trends act in opposite directions with respect to product selectivity, their combined effect allows to maintain favorable yields in the kerosene range. 

The results presented in \autoref{Fig.: Operating parameters results} demonstrate the substantial variability in optimal process configurations, not only with respect to the selected processes and their installed capacities, but also in terms of the operation of those processes. These variations are strongly influenced by external parameters, such as electricity pricing and \ce{CO2} emission constraints. It is evident that the process set points, including temperature, pressure, and inlet composition, must be adapted to prevailing techno-economic conditions to ensure cost-effective system performance. In contrast, constraining these operational variables to static values, independent of external influences, can result in suboptimal plant performance and increased specific production costs, as quantified in \autoref{Fig.: Results fixed vs. adaptable design}.

\subsection{Heat integration}
\begin{figure}[h]
\centering
\includegraphics[width=0.6\columnwidth]{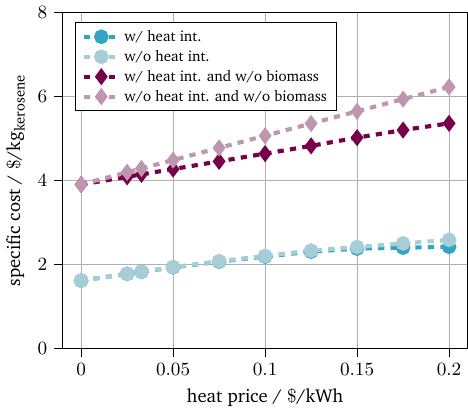}
\caption{Results of the sensitivity analysis for the specific kerosene production costs as a function of the heat price. The blue curves represent the kerosene costs, once with (w/ heat int.) and once without heat integration (w/o heat int.). The red lines show the same trends, but excluding the use of biomass gasification (w/o biomass) and \ce{CO2} from point sources. All results are generated excluding the autothermal reforming (ATR) process and there are no restrictions regarding \ce{CO2} emissions.}
\label{Fig.: Heat integration results}
\end{figure}

\autoref{Subsubsection: Heat integration} presents an advanced formulation for process level heat integration, in which discrete thermal sources and sinks are defined for each process within the overall synthesis pathway. To quantitatively assess the influence of heat integration on the specific production costs of sustainable kerosene, a parametric sensitivity analysis is conducted with respect to the unit cost of externally supplied thermal energy. The results are depicted in \autoref{Fig.: Heat integration results}, wherein the external heat price is varied between \SI{0}{\$/kWh} and \SI{0.2}{\$/kWh}.
All scenarios exclude the ATR process and impose no constraints on \ce{CO2} emissions. The blue data points represent cost for system configurations both with (dark blue) and without (light blue) the implementation of heat integration. Costs shown in red correspond to designs that omit the use of biomass as a feedstock and exclude \ce{CO2} from industrial point sources (“w/o biomass”).

Analyzing the costs represented by the blue curve including heat integration, the specific kerosene production costs range from \SI{1.61}{\$/kg\textsubscript{kerosene}} at a heat price of \SI{0}{\$/kWh} to \SI{2.42}{\$/kg\textsubscript{kerosene}} at \SI{0.2}{\$/kWh}, corresponding to an approximate \SI{50}{\%} increase in production costs attributable to the rising cost of externally supplied thermal energy. The light blue curve, which reflects system configurations without thermal integration, follows a comparable trend. At the upper bound of heat prices, the specific cost increases to \SI{2.58}{\$/kg\textsubscript{kerosene}}, indicating that heat integration achieves a marginal cost reduction of only \SI{7}{\%}. This limited cost differential is primarily attributable to the underlying plant design. In both cases, synthesis gas is produced \emph{via} biomass gasification. Given the constraints imposed by the minimum temperature approach for heat exchange, the only significant thermal integration potential lies between the initial endothermic decomposition step of biomass and the subsequent exothermic gasification step, provided that the gasifier operates under conditions conducive to net heat release. In the absence of heat integration, the sole deviation is the need for external thermal energy input to support biomass decomposition, leading to a modest rise in overall production costs. Thus, under these specific design and operational conditions, the economic incentive for implementing thermal integration remains limited.

In contrast, when the utilization of biomass, point source-derived \ce{CO2}, and the ATR process are entirely excluded from the system configuration, the significance of thermal integration becomes markedly pronounced. Under these constraints, all carbon required for FT synthesis must be supplied \emph{via} atmospheric \ce{CO2}, necessitating the deployment of a DAC process. The DAC system imposes a substantial thermal demand, particularly for the desorption phase, which operates at approximately \SI{100}{\degreeCelsius}. The inherently exothermic nature of the FT synthesis reaction presents a favorable opportunity for thermal integration, enabling recovery and utilization of waste heat to satisfy the DAC heat requirements. In scenarios where heat integration is not considered, the thermal energy demand for \ce{CO2} capture must be fulfilled through external heat supply. A comparative analysis of the two extreme cases, at a heat price of \SI{0.2}{\$/kWh}, reveals that the specific kerosene production cost increases from \SI{5.35}{\$/kg\textsubscript{kerosene}} (with heat integration) to \SI{6.22}{\$/kg\textsubscript{kerosene}} (without integration), corresponding to an approximate \SI{14}{\%} cost reduction due to internal heat integration. More broadly, the magnitude of cost deviation between the heat-integrated and non-integrated configurations is consistently greater for the red curves compared to the blue curves, indicating a stronger dependence on thermal integration in systems relying solely on atmospheric \ce{CO2} as the carbon source. This underscores the conclusion that while the benefit of heat integration is design-dependent, it becomes critically important in process configurations characterized by high thermal demand and limited internal heat sources.

\section*{Conclusions and outlook}
\phantomsection
\addcontentsline{toc}{section}{Conclusions and outlook} 
\label{Section: Conclusions and outlook}
This work introduces an advanced multi-objective superstructure optimization framework considering component mixtures, that integrates ANNs within a MIQCP formulation. By modeling mixtures and embedding data-driven surrogate models into the mathematical optimization structure, the proposed methodology overcomes the inherent limitations of conventional superstructure-based approaches reported in the literature. Specifically, the framework enables the consideration of variable input and output stream compositions, thereby allowing for the simultaneous optimization of process configurations and target product composition. Moreover, the formulation captures complex, nonlinear input–output relationships of individual process units and facilitates the joint optimization of both discrete system-level decisions (\emph{e.g.}\@, process selection and capacity allocation) and continuous process operating parameters (\emph{e.g.}\@, pressure, temperature, and feed composition).

The optimization results indicate that, in the absence of \ce{CO2} emission constraints, least-cost process configurations are characterized by the exclusive deployment of the fossil-based ATR route. This yields low specific production costs for kerosene (\textapprox\SI{0.79}{\$/kg\textsubscript{kerosene}}), but is associated with elevated specific greenhouse gas emissions on the order of \SI{4.47}{kg\textsubscript{\ce{CO2}}/kg\textsubscript{kerosene}}. Constraints on carbon emissions necessitate the integration of biomass gasification in conjunction with \ce{CO2} removal \emph{via} DAC-CS. These system configurations result in significantly higher production costs exceeding \SI{2.3}{\$/kg\textsubscript{kerosene}}, while simultaneously achieving substantial reductions in net \ce{CO2} emissions. When supply chain emissions are taken into account, achieving net zero \ce{CO2} emissions is infeasible through biomass utilization alone and carbon sequestration is required.

Furthermore, \ce{CO2} abatement costs (\textapprox\SI{390}{\$/t\textsubscript{\ce{CO2}}}) align well with literature values, indicating that high carbon pricing would be required to render bio-based SAF production economically competitive. The analysis highlights that mixed feedstock approaches (\emph{e.g.}\@, combining natural gas and pine chips) offer optimal trade offs between costs and emissions (\textapprox\SI{2.38}{\$/kg\textsubscript{kerosene}} at near zero \ce{CO2}). At zero emissions, a bio-based system is slightly less costly than a combination of synthesis from natural gas and offsets \emph{via} DAC-CS, while a system reliant on atmospheric carbon is around five times as expensive. 

Crucially, the study underlines the value of simultaneous optimization of the overall process system topology and individual process operating parameters. Designs that leverage flexible process parameters -- enabled by embedded ANNs -- consistently outperform fixed configurations, achieving up to \SI{20}{\%} cost savings. Sensitivity studies show how operating conditions such as FT reactor pressure and gasification temperature respond to external factors such as electricity prices.

Several limitations inherent to the present study warrant further investigation in subsequent research. Primarily, the optimization of the ANN input and output variables is contingent upon the accuracy and fidelity of the underlying process simulations performed in Aspen Plus\textsuperscript{\textregistered} \cite{AspenPlu.24.08.2023}. In particular, the FT reactor is modeled using an empirical surrogate, which could be substituted with a more mechanistically rigorous model to enhance the predictive accuracy of product distribution as a function of key operating parameters. Moreover, only three subprocesses within the superstructure are represented \emph{via} embedded ANN surrogates. This restriction arises from the increase in computational complexity associated with the inclusion of additional binary decision variables, particularly those introduced by the ReLU ANNs. Future research should aim to improve the computational tractability of the mixed-integer optimization problem, thereby enabling the integration of a larger number of data-driven surrogate models. From a process systems engineering standpoint, the scope of this study is limited to the production of the kerosene fraction of the FT product stream, with a specific focus on linear alkanes (n-alkanes) as the primary target compounds. For a more comprehensive understanding of SAF production pathways, downstream upgrading processes (such as hydrocracking and hydrogenation) should be incorporated. Additionally, alternative synthesis routes, such as the methanol-to-jet pathway, could be explored. This would also enable consideration of a more diverse set of hydrocarbon products, including iso-alkanes and cyclo-alkanes, yielding a product composition that more closely aligns with fuel specifications.

\section*{Author contributions}
\textbf{\href{https://scholar.google.com/citations?user=aERoihcAAAAJ&hl=de&oi=ao}{Alexander Klimek}}: conceptualization, methodology, software, data curation, original draft; \textbf{\href{https://www.mathopt.de/people/plate/}{Christoph Plate}}: methodology, software; \textbf{\href{https://scholar.google.com/citations?user=_-2c3bQAAAAJ&hl=de&oi=ao}{Sebastian Sager}}: methodology, supervision, resources; \textbf{\href{https://scholar.google.com/citations?user=MNTZeJIAAAAJ&hl=de&oi=ao}{Kai Sundmacher}}: supervision, resources, review \& editing; \textbf{\href{https://scholar.google.com/citations?user=g7djIaMAAAAJ&hl=en}{Caroline Ganzer}}: conceptualization, methodology, supervision, review \& editing

\section*{Conflicts of interest}
There are no conflicts to declare.

\section*{Data availability}
The data supporting this article have been included as part of the Supplementary Information.

\section*{Acknowledgements}
We acknowledge financial support from the research initiative "SmartProSys: Intelligent Process Systems for the Sustainable Production of Chemicals" funded by the Ministry for Science, Energy, Climate Protection and the Environment of the State of Saxony-Anhalt, from the European Regional Development Fund (grant "Weiterentwicklung des Center for Dynamic Systems, CDS") under the European Union's Horizon Europe Research and Innovation Program, and from the German Research Foundation (DFG) within the priority program 2331 "Machine Learning in Chemical Engineering" under grant SA 2016/3-1. 

The authors would like to thank \textbf{\href{https://scholar.google.com/citations?user=jpg9q30AAAAJ&hl=en}{Manuel Garcia-Perez}} for his helpful inputs regarding the modeling of gasification processes.

OpenAI's ChatGPT was utilized to aid the development of code and to improve the writing of this manuscript. All outputs were carefully reviewed and edited by the authors to ensure accuracy and scientific rigor.

\renewcommand\refname{References}

\bibliography{Literature} 

\newcounter{mainLastPage}
\setcounter{mainLastPage}{\value{page}}

\clearpage
\pagestyle{empty}

\includepdf[pages=-,pagecommand={}]{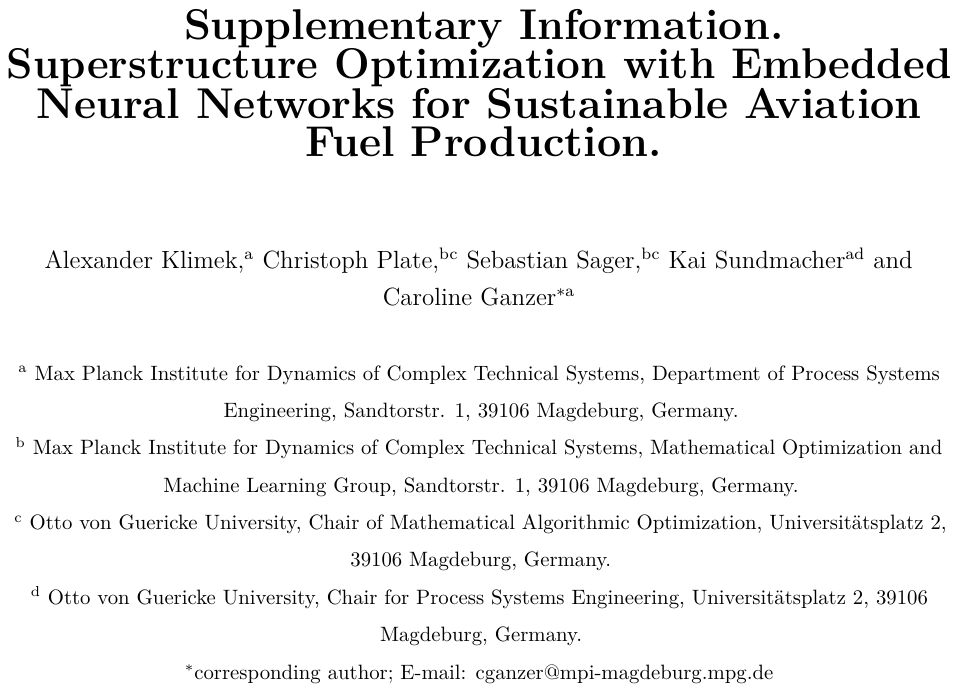}

\end{document}